\documentclass[preprint]{ptephy_v1}%%%%%% to generate preprint number
\usepackage{amsmath} %for dealing with mathematics,
\usepackage{amsthm} %for dealing with theorem environments,
\usepackage{hyperref} %for linking the cross references
\usepackage{graphics} %for dealing with figures.
\usepackage{algorithmic} %for describing algorithms
\usepackage{subfig} %for getting the subfigures e.g., "Figure 1a and 1b" etc.
\usepackage{url} %It provides better support for handling andbreaking URLs.
\usepackage{grffile}
\usepackage{color}
\begin{document}
\title{Search for Tens of MeV Neutrinos associated with Gamma-Ray Bursts in Super-Kamiokande}

%%%% To generate auto affiliation numbers please use \author{}\affil{} command
\newcommand{\AFFicrr}{1}
\newcommand{\AFFkashiwa}{2}
\newcommand{\AFFicrrkashiwa}{3}
\newcommand{\AFFmad}{4}
\newcommand{\AFFbu}{5}
\newcommand{\AFFbcit}{6}
\newcommand{\AFFuci}{7}
\newcommand{\AFFcsu}{8}
\newcommand{\AFFcnm}{9}
\newcommand{\AFFduke}{10}
\newcommand{\AFFllr}{11}
\newcommand{\AFFfukuoka}{12}
\newcommand{\AFFgifu}{13}
\newcommand{\AFFgist}{14}
\newcommand{\AFFuh}{15}
\newcommand{\AFFicl}{16}
\newcommand{\AFFbari}{17}
\newcommand{\AFFnapoli}{18}
\newcommand{\AFFpadova}{19}
\newcommand{\AFFroma}{20}
\newcommand{\AFFkeio}{21}
\newcommand{\AFFkek}{22}
\newcommand{\AFFkcl}{23}
\newcommand{\AFFkobe}{24}
\newcommand{\AFFkyoto}{25}
\newcommand{\AFFliv}{26}
\newcommand{\AFFmiyagi}{27}
\newcommand{\AFFnagoya}{28}
\newcommand{\AFFkmi}{29}
\newcommand{\AFFpol}{30}
\newcommand{\AFFsuny}{31}
\newcommand{\AFFokayama}{32}
\newcommand{\AFFosaka}{33}
\newcommand{\AFFox}{34}
\newcommand{\AFFqmul}{35}
\newcommand{\AFFral}{36}
\newcommand{\AFFseoul}{37}
\newcommand{\AFFsheff}{38}
\newcommand{\AFFshizuokasc}{39}
\newcommand{\AFFstfc}{40}
\newcommand{\AFFskk}{41}
\newcommand{\AFFtokai}{42}
\newcommand{\AFFtokyo}{43}
\newcommand{\AFFtodai}{44}
\newcommand{\AFFipmu}{45}
\newcommand{\AFFtit}{46}
\newcommand{\AFFtus}{47}
\newcommand{\AFFtoronto}{48}
\newcommand{\AFFtriumf}{49}
\newcommand{\AFFtsinghua}{50}
\newcommand{\AFFwu}{51}
\newcommand{\AFFwarwick}{52}
\newcommand{\AFFwinnipeg}{53}
\newcommand{\AFFynu}{54}

\author{
%%%%%%%%%%%%%%%%%%%%%%%%%%%%%%%%%%%%%%%%%%%%%%%%%%%%%%%%%%%%%%%%%%%%
%ICRR
\name{A.~Orii}{\AFFicrr},
\name{K.~Abe}{\AFFicrr,\AFFipmu},
\name{C.~Bronner}{\AFFicrr},
\name{Y.~Hayato}{\AFFicrr,\AFFipmu},
\name{M.~Ikeda}{\AFFicrr},
\name{S.~Imaizumi}{\AFFicrr},
\name{H.~Ito}{\AFFicrr}, 
\name{J.~Kameda}{\AFFicrr,\AFFipmu},
\name{Y.~Kataoka}{\AFFicrr},
\name{Y.~Kato}{\AFFicrr}
% ...
%\name{(The Super-Kamiokande Collaboration)}{}}
\collaborator{(The Super-Kamiokande Collaboration)}}
%%%%%%%%%%%%%%%%%%%%%%%%%%%%%%%%%%%%%%%%%%%%%%%%%%%%%%%%%%%%%%%%%%%%%
%\address{
\affil[\AFFicrr]{{Kamioka Observatory, Institute for Cosmic Ray Research, University of Tokyo, Kamioka, Gifu 506-1205, Japan}}
\affil[\AFFkashiwa]{{Research Center for Cosmic Neutrinos, Institute for Cosmic Ray Research, University of Tokyo, Kashiwa, Chiba 277-8582, Japan}}
\affil[\AFFicrrkashiwa]{{Institute for Cosmic Ray Research, University of Tokyo, Kashiwa, Chiba 277-8582, Japan}}
\affil[\AFFmad]{{Department of Theoretical Physics, University Autonoma Madrid, 28049 Madrid, Spain}}
\affil[\AFFbu]{{Department of Physics, Boston University, Boston, MA 02215, USA}}
\affil[\AFFbcit]{{Department of Physics, British Columbia Institute of Technology, Burnaby, BC, V5G 3H2, Canada }}
\affil[\AFFuci]{{Department of Physics and Astronomy, University of California, Irvine, Irvine, CA 92697-4575, USA }}
\affil[\AFFcsu]{{Department of Physics, California State University, Dominguez Hills, Carson, CA 90747, USA}}
\affil[\AFFcnm]{{Institute for Universe and Elementary Particles, Chonnam National University, Gwangju 61186, Korea}}
\affil[\AFFduke]{{Department of Physics, Duke University, Durham NC 27708, USA}}
\affil[\AFFllr]{{Ecole Polytechnique, IN2P3-CNRS, Laboratoire Leprince-Ringuet, F-91120 Palaiseau, France }}
\affil[\AFFfukuoka]{{Junior College, Fukuoka Institute of Technology, Fukuoka, Fukuoka 811-0295, Japan}}
\affil[\AFFgifu]{{Department of Physics, Gifu University, Gifu, Gifu 501-1193, Japan}}
\affil[\AFFgist]{{GIST College, Gwangju Institute of Science and Technology, Gwangju 500-712, Korea}}
\affil[\AFFuh]{{Department of Physics and Astronomy, University of Hawaii, Honolulu, HI 96822, USA}}
\affil[\AFFicl]{{Department of Physics, Imperial College London , London, SW7 2AZ, United Kingdom }}
\affil[\AFFbari]{{ Dipartimento Interuniversitario di Fisica, INFN Sezione di Bari and Universit\`a e Politecnico di Bari, I-70125, Bari, Italy}}
\affil[\AFFnapoli]{{Dipartimento di Fisica, INFN Sezione di Napoli and Universit\`a di Napoli, I-80126, Napoli, Italy}}
\affil[\AFFpadova]{{Dipartimento di Fisica, INFN Sezione di Padova and Universit\`a di Padova, I-35131, Padova, Italy}}
\affil[\AFFroma]{{INFN Sezione di Roma and Universit\`a di Roma ``La Sapienza'', I-00185, Roma, Italy}}
\affil[\AFFkeio]{{Department of Physics, Keio University, Yokohama, Kanagawa, 223-8522, Japan}}
\affil[\AFFkek]{{High Energy Accelerator Research Organization (KEK), Tsukuba, Ibaraki 305-0801, Japan }}
\affil[\AFFkcl]{{Department of Physics, King's College London, London, WC2R 2LS, UK }}
\affil[\AFFkobe]{{Department of Physics, Kobe University, Kobe, Hyogo 657-8501, Japan}}
\affil[\AFFkyoto]{{Department of Physics, Kyoto University, Kyoto, Kyoto 606-8502, Japan}}
\affil[\AFFliv]{{Department of Physics, University of Liverpool, Liverpool, L69 7ZE, United Kingdom}}
\affil[\AFFmiyagi]{{Department of Physics, Miyagi University of Education, Sendai, Miyagi 980-0845, Japan}}
\affil[\AFFnagoya]{{Institute for Space-Earth Environmental Research, Nagoya University, Nagoya, Aichi 464-8602, Japan}}
\affil[\AFFkmi]{{Kobayashi-Maskawa Institute for the Origin of Particles and the Universe, Nagoya University, Nagoya, Aichi 464-8602, Japan}}
\affil[\AFFpol]{{National Centre For Nuclear Research, 02-093 Warsaw, Poland}}
\affil[\AFFsuny]{{Department of Physics and Astronomy, State University of New York at Stony Brook, NY 11794-3800, USA}}
\affil[\AFFokayama]{{Department of Physics, Okayama University, Okayama, Okayama 700-8530, Japan }}
\affil[\AFFosaka]{{Department of Physics, Osaka University, Toyonaka, Osaka 560-0043, Japan}}
\affil[\AFFox]{{Department of Physics, Oxford University, Oxford, OX1 3PU, United Kingdom}}
\affil[\AFFqmul]{{School of Physics and Astronomy, Queen Mary University of London, London, E1 4NS, United Kingdom}}
\affil[\AFFral]{{Rutherford Appleton Laboratory, Harwell, Oxford, OX11 0QX, UK }}
\affil[\AFFseoul]{{Department of Physics, Seoul National University, Seoul 151-742, Korea}}
\affil[\AFFsheff]{{Department of Physics and Astronomy, University of Sheffield, S3 7RH, Sheffield, United Kingdom}}
\affil[\AFFshizuokasc]{{Department of Informatics in Social Welfare, Shizuoka University of Welfare, Yaizu, Shizuoka, 425-8611, Japan}}
\affil[\AFFstfc]{{STFC, Rutherford Appleton Laboratory, Harwell Oxford, and Daresbury Laboratory, Warrington, OX11 0QX, United Kingdom}}
\affil[\AFFskk]{{Department of Physics, Sungkyunkwan University, Suwon 440-746, Korea}}
\affil[\AFFtokai]{{Department of Physics, Tokai University, Hiratsuka, Kanagawa 259-1292, Japan}}
\affil[\AFFtokyo]{{The University of Tokyo, Bunkyo, Tokyo 113-0033, Japan }}
\affil[\AFFtodai]{{Department of Physics, University of Tokyo, Bunkyo, Tokyo 113-0033, Japan }}
\affil[\AFFipmu]{{Kavli Institute for the Physics and Mathematics of the Universe (WPI), The University of Tokyo Institutes for Advanced Study, University of Tokyo, Kashiwa, Chiba 277-8583, Japan }}
\affil[\AFFtit]{{Department of Physics,Tokyo Institute of Technology, Meguro, Tokyo 152-8551, Japan }}
\affil[\AFFtus]{{Department of Physics, Faculty of Science and Technology, Tokyo University of Science, Noda, Chiba 278-8510, Japan }}
\affil[\AFFtoronto]{{Department of Physics, University of Toronto, ON, M5S 1A7, Canada }}
\affil[\AFFtriumf]{{TRIUMF, 4004 Wesbrook Mall, Vancouver, BC, V6T2A3, Canada }}
\affil[\AFFtsinghua]{{Department of Engineering Physics, Tsinghua University, Beijing, 100084, China}}
\affil[\AFFwu]{{Faculty of Physics, University of Warsaw, Warsaw, 02-093, Poland }}
\affil[\AFFwarwick]{{Department of Physics, University of Warwick, Coventry, CV4 7AL, UK }}
\affil[\AFFwinnipeg]{{Department of Physics, University of Winnipeg, MB R3J 3L8, Canada }}
\affil[\AFFynu]{{Department of Physics, Yokohama National University, Yokohama, Kanagawa, 240-8501, Japan}}

%%% To include the collaborator name... Please use the command "\collaborator"
%%% For example: \collaborator{ATLAS Collaboration}
%\collaborator{SK Collaboration}

\begin{abstract}
A search for neutrinos produced in coincidence with Gamma-Ray Bursts (GRBs) was conducted with the Super-Kamiokande (SK) detector. Between December 2008 and March 2017, the Gamma-ray Coordinates Network recorded 2208 GRBs that occurred during normal SK operation. Several time windows around each GRB were used to search for coincident neutrino events. No statistically significant signal in excess of the estimated backgrounds was detected. The $\bar\nu_e$ fluence in the range from 8~MeV to 100~MeV in positron total energy for $\bar\nu_e+p\rightarrow e^{+}+n$ was found to be less than $\rm 5.07\times10^5$~cm$^{-2}$ per GRB in 90\% C.L. For all GRBs, upper bounds were obtained on the fluence as a function of neutrino energy. Additionally, for GRBs at known distances, upper limits were set for the neutrino energy emission at the GRB.
\end{abstract}

\maketitle
\section{Introduction}
Gamma-Ray Bursts (GRBs) are one of the most luminous phenomena in the universe, as an enormous amount of energy is released as gamma-rays over a very short time scale. GRBs were first discovered in 1967~\cite{GRB1973}; since that time, detailed observations have revealed their main features, though the underlying astrophysical mechanism remains poorly understood. Neutrinos may be able to advance our understanding of the burst mechanism, as they interact very weakly with matter and, thus, can promptly escape from high-density regions such as the inner core of a GRB progenitor. The observation of neutrinos from a GRB would enable a direct study of the entire burst process.

GRBs can be sub-divided into two categories, based on their duration. Long GRBs are believed to originate when the rotating core of a massive star collapses into a neutron star (NS) or a black hole (BH). Short GRBs may arise from a NS-BH merger, or from the merger of two NS~\cite{Meszaros2006}.
Either in long or short GRBs, gamma-ray emission is thought to be produced by relativistic outflow. The central engine of GRBs may also emit a large amount of thermal neutrinos. A typical total energy of neutrinos is expected to be $E_{\nu,tot} \sim 2 \times 10^{53}$ erg~\cite{lownu4}, 
which gives a fluence at the Earth to be
\begin{equation}
\Phi_{\nu}=1\times\left(\frac{E_{\nu,tot}}{2\times10^{53}\,\rm{erg}}\right)\left(\frac{65\,\rm{MeV}}{E_{\nu}}\right)\left(\frac{4000\,\rm{Mpc}}{D} \right)^2 \rm{cm}^{-2}
\end{equation}
where $E_{\nu,tot}$ is the total neutrino energy of the fireball, $E_{\nu}$ is average neutrino energy at the source, and $D$ is the distance to the GRB.

For a GRB of typical energy at redshift $z$=2, the expected fluence is $\Phi_\nu \sim 1\,\rm{cm}^{-2}$ for neutrinos of $\mathcal{O}(10)$~MeV. A previous study with the Super-Kamiokande (SK) detector established an upper limit on the fluence of MeV-scale neutrinos of $10^5-10^8$~$\rm{cm}^{-2}$~\cite{SK2002}, which is significantly larger than the expected fluence.
However, predictions of neutrino fluence are highly model-dependent. For instance, a cosmic string model~\cite{lownu2}, which uses new physics predicts a fluence of $\Phi_{\nu} \sim 10^7-10^8$~$\rm{cm}^{-2}$~\cite{lownu4}, which is many orders of magnitude higher than that expected from the fireball scenarios.

Searches for neutrinos associated with GRBs have been conducted by SK~\cite{SK2002, SK2009}, KamLAND~\cite{Kamland2015}, Borexino~\cite{Borexino2016}, IceCube~\cite{IceCube}, ANITA~\cite{ANITA2011}, ANTARES~\cite{ANTARES2013}, SNO~\cite{SNO2014} and BUST~\cite{BUST2015}.
As yet, no clear evidence has been observed for a neutrino signal originating from a GRB.
This paper describes a new search with an increased number of GRBs and a longer exposure of SK data.

This paper is organized as follows. 
In Section 2, we describe the search method including the GRB catalog, the SK detector, data reduction, and analysis methods.
In Section 3, we present the results of searches. 
Finally, in Section 4, we conclude the summary of the results.

\section{Search Method}
In order to search for a possible GRB neutrino signal, we looked for correlations between GRBs in the Gamma-ray 
Coordinates Network (GCN) database~\cite{GCN} and SK data taken from 2008 December 7 to 2017 May 31. We have adopted the GRBweb online catalog~\cite{IClist}, which was developed for neutrino searches with the IceCube observatory. The catalog is compiled from GCN circulars that archive satellite reports. It contains important information about GRB events, including duration ($t90$), distance from Earth, trigger time ($t0$), start time ($t1$) and end time ($t2$). To avoid confusion with other variables, this paper will refer to the start and end times as $t_s$ and $t_e$. GRB events with a $t90$ duration less than 2 seconds are defined as `short GRBs', while GRBs with greater duration are classified as `long GRBs'.

The catalog contains 2208 GRB events observed during the SK data-taking period considered for this analysis. However, fourteen of these GRBs are missing either $t_s$ or $t_e$ information and the remaining sample of 2194~GRBs is considered for variable timing window analysis discussed in Sec.~\ref{sec:analysis_t1t2}.

\subsection{SK detector}
Super-Kamiokande is a 50 kton water Cherenkov detector located in the Kamioka mine~\cite{SKdetector}. A rock overburden of 1000 meters (2700 meters water equivalent) reduces the flux of cosmic ray muons by five orders of magnitude. The SK detector is a right cylinder with a diameter of 39.3 m and height of 41.4 m. A stainless steel frame divides the detector into two optically separated volumes, an inner detector (ID) and outer detector (OD). The steel frame provides both the optical barrier and a support structure for photo-multiplier tubes (PMTs). 

The OD serves as both a passive shield and active veto for external particles. It contains a 2.5~m thick layer of water and is instrumented with 1885 8~inch PMTs. The water stops gamma rays from the surrounding rock, while the PMTs enable the OD to tag particles originating from outside of the detector.

The ID contains 32.5 kton of water, which is viewed by 11,129 20~inch PMTs for a total photocathode coverage of 40\%. In areas that are close to the ID wall, the background rate is very high. For this analysis, a 22.5 kton fiducial mass is defined as the water contained in a virtual volume that is at least 2~m from the ID wall. This region of the detector is called the fiducial volume (FV).

\subsection{Data reduction}\label{sec:reduction}
Super-Kamiokande has performed two different analyses for neutrinos with energies below 100~MeV. The solar neutrino analysis~\cite{solar} focuses on high levels of background reduction to enable precision measurements of $^8$B solar neutrinos. The supernova relic neutrino (SRN) analysis~\cite{relic2} is optimized to search for electron antineutrinos interacting via the inverse beta decay (IBD) reaction ($\bar\nu_{e}+p\rightarrow e^{+}+n $).

In the energy region of this analysis (8~MeV to 100~MeV in positron total energy), 
the dominant reaction is the IBD reaction~\cite{SVcalc} assuming that 
the neutrino flux from a GRB is distributed approximately equally among all neutrino flavors; 
therefore, this study adopts the data reduction of the SRN analysis. 
Expected signals are single particle electron-like events.  
A detailed description of the event reconstruction and the data reduction can be found in~\cite{relic2}. 
The selection cuts below were tuned following an upgrade to the detector electronics in 2008. First, non-physical events are removed, as well as decay electrons identified with cosmic ray muons, and events near the ID wall; a summary of the selection is listed here:

\begin{itemize}
\item{\bf Calibration event cut:}
  Periodic calibration events induced by laser and Xe light during normal data taking are identified by a trigger tag and removed.
\item{\bf Noise event cut:}
  Events caused by electronic noise are removed.
\item{\bf OD cut:}
  Charged particles originating from outside of the detector are identified by their signal in the OD and removed.
\item{\bf Time difference cut:}
  Decay electrons and noise events produced by PMT ringing after cosmic ray muons are removed via a 50~$\rm \mu$s cut following each event.
\item{\bf Fiducial volume cut:}
  Radioactive backgrounds emanating from PMTs and the steel frame are identified by their proximity to the ID wall and removed.
\item{\bf Goodness cut:}
  The energy, vertex and direction of neutrino candidate events are reconstructed according to the method described in~\cite{solar}. Candidate events that are reconstructed poorly are removed.
\item{\bf Energy cut:}
  Neutrino candidate events with reconstructed energy below 8~MeV are outside the range of this analysis and are therefore removed.
\end{itemize}

After this sequence of data reduction (called `first reduction' hereafter), the dominant background source in the energy range below 20~MeV is the radioactive decay of unstable nuclei. These nuclei are created by spallation-induced by cosmic ray muons. In the energy range above 20~MeV, the dominant background originates from atmospheric neutrino interactions, which can take the form of pions, muons, gamma-rays (via neutral current interactions), and decay electrons produced by the decay of muons with energies below the Cherenkov threshold.

To remove these backgrounds, additional selection criteria are applied. These are described below:

\begin{figure}[htp]
  \begin{center}
    \includegraphics[width=6cm]{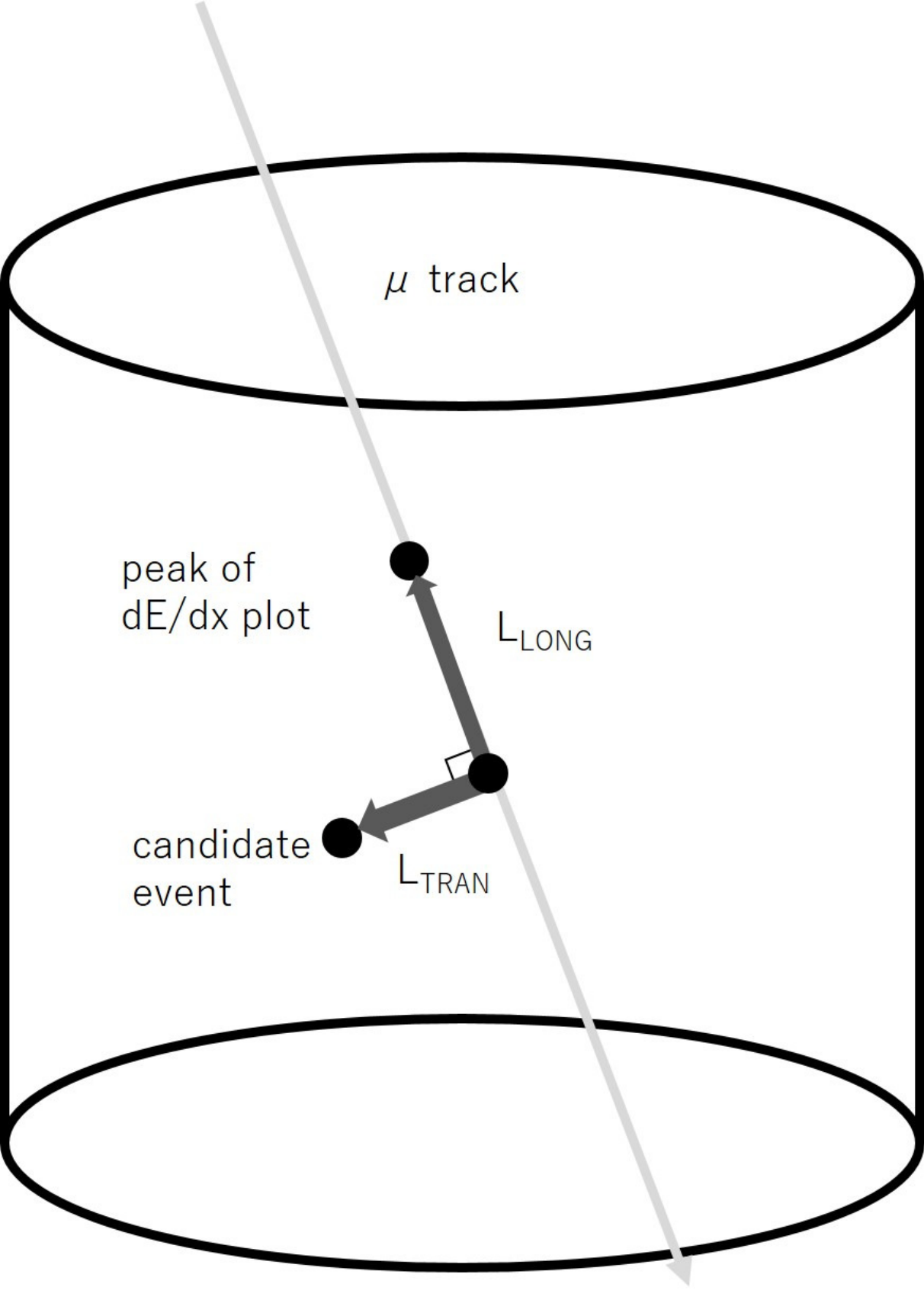}
    \caption{Schematic showing spallation distance variables. $L_{LONG}$ is the longitudinal distance from the spallation point to where energy deposit of the muon is maximum. $L_{TRAN}$ is the transverse distance from the muon track to the neutrino candidate event.}
    \label{fig:spa_sch}
  \end{center}
\end{figure}

\begin{itemize}
\item{\bf Spallation cut:}
  The removal of spallation products relies on several factors. Both the time difference to a muon event, and the transverse distance from a muon track are considered. Additionally, the maximum value of energy deposited by the muon in a 50~cm bin along muon track, and the longitudinal distance from the spallation point where energy deposit of the muon is maximum (Fig.~\ref{fig:spa_sch}) are used. To identify spallation-induced events in the detector, these variables are calculated for muons observed within a 30~sec window prior to the neutrino candidates.
\item{\bf Gamma cut:}
  A more stringent cut of events originating from outside of the detector is implemented by using the reconstructed travel distance from the ID wall. It is calculated from the vertex position and the event direction.
\item{\bf Pion cut:}
  The sharpness of the Cherenkov ring is evaluated and used to remove charged pions created in atmospheric neutrino interactions.  
\item{\bf OD correlated event cut:}
  Incoming events without an explicit OD trigger are identified and removed by searching for correlations between the ID hits and the OD hits.
\item{\bf Multi-ring cut:}
  Atmospheric neutrino interactions can sometimes produce both a charged lepton and a charged pion; such events have two Cherenkov rings. To reduce this background, events with an angle between rings that is greater than 60 degrees are removed from the data sample.
\item{\bf Solar events cut:}
  The direction to the sun, as well as total energy and anisotropy of the PMT hit patterns is used to remove solar neutrinos from the data sample.  
\item{\bf Pre/Post activity cut:}
Events with $>$ 12 hits for pre-activity or $>$ 15 hits  for post-activity 
within the time range from -5 $\mu$sec to +35 $\mu$sec are removed. Such activities are expected in low energy atmospheric interactions which produce muons at or below the Cherenkov threshold~\cite{relic2}. 

\item{\bf Cherenkov angle cut:}
  Low energy atmospheric muon neutrinos can produce low energy muons via charged current interactions. These muons will be close to the Cherenkov threshold and, thus, have a smaller Cherenkov angle than that of a highly relativistic charged particle. Neutral current interactions and nuclear de-excitations can produce events comprising multiple gamma-rays. These reconstruct as if they had a large Cherenkov angle, due to an artifact of the fitting algorithm. By placing a constraint of 38 degrees to 50 degrees on the reconstructed Cherenkov angle, both of these backgrounds can be removed.
\item{\bf $\rm \mu/\pi$ cut:}
  Residual muon and pion events remaining after the pion cut and the Cherenkov angle cut are removed by using the ratio of PMTs with large charge. These events deposit a large amount of energy along a short track, causing one PMT to observe many photoelectrons.
\item{\bf N16 cut:}
Low energy muons may be captured by $\rm ^{16}$O to produce $\rm ^{16}$N. The $\rm ^{16}$N will decay by emitting a $\rm \gamma$ and/or an electron. Spatial and timing correlations between low energy events and stopping muons are used to remove this background.
\end{itemize}

\begin{figure}[htp]
  \begin{center}
    \includegraphics[width=8cm]{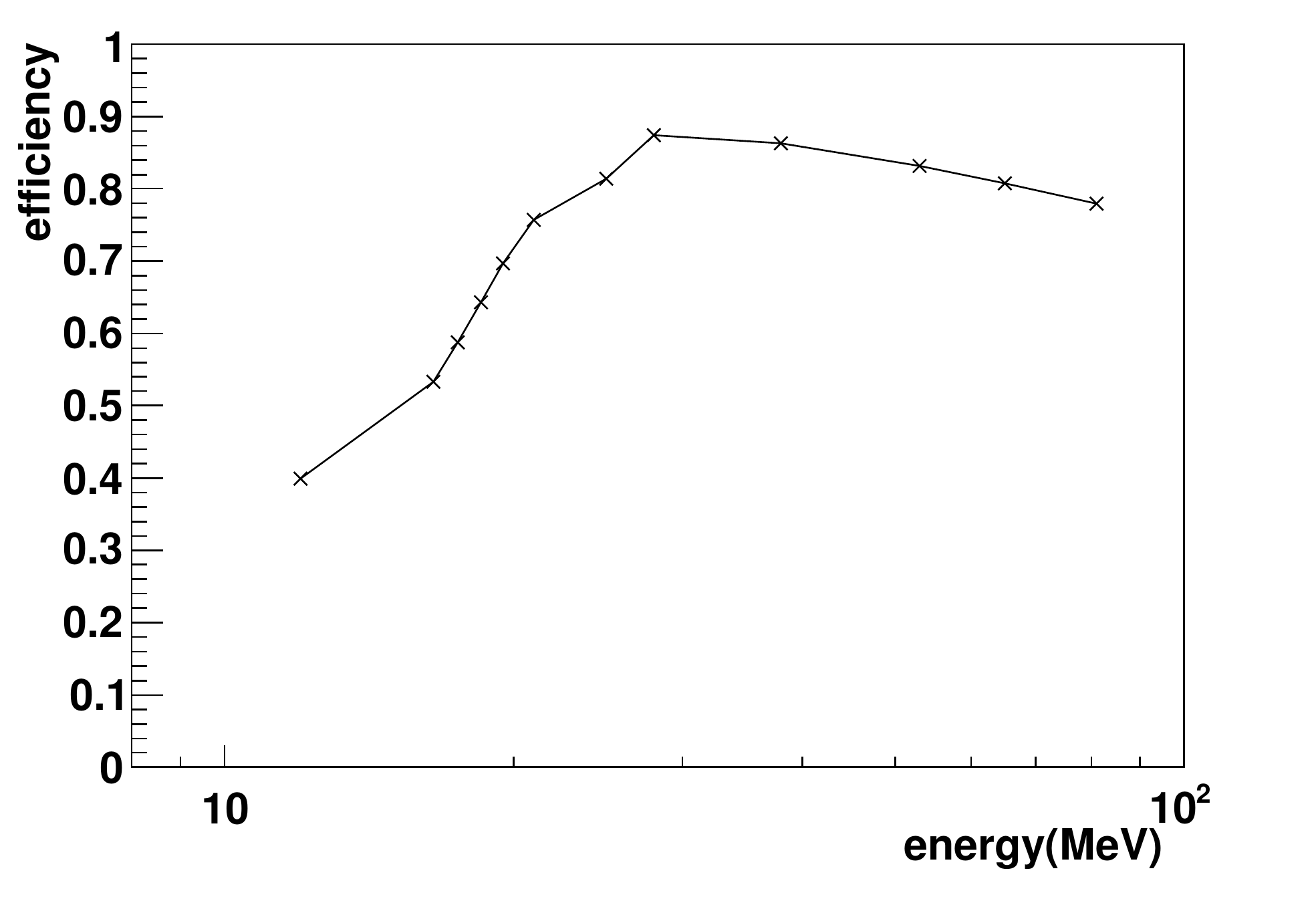}
    \caption{Total signal efficiency versus generated positron energy. 
     Below 30~MeV, the efficiency increases with energy because signal 
     loss due to the solar event cut and the spallation cut are reduced.
      Above 30~MeV, the efficiency begins to decrease as the pion cut and the $\mu/\pi$ cut remove high energy events.}
    \label{fig:eff}
  \end{center}
\end{figure}

The method described in~\cite{relic2} is used to determine the efficiency of the signal retained by the spallation cut, with the resulting number given in~\cite{Orii_thesis}. 
The signal efficiency for the other reduction steps described above are calculated using a Monte Carlo simulation of events generated at typical energies within the range of this analysis. The generated events are positrons distributing uniformly in the 32~kton ID in isotropic direction. By combining these procedures, we obtain the total signal efficiency as a function of positron energy in the fiducial volume, which is shown in Fig.~\ref{fig:eff}.

\subsection{Neutrino search for individual GRBs}
Two different timing window configurations were used to search for an excess of neutrino events correlated with individual GRB timings. In Sec.~\ref{sec:analysis_500_sec}, we describe an analysis with a fixed time window of 
$\pm$500~sec around the GRB trigger time $t0$. The analysis described in Sec.~\ref{sec:analysis_t1t2} uses the GRB start and end times to determine the duration of a variable timing window. The results of these analyses are presented in Sec.~\ref{sec:results_500_sec} and Sec.~\ref{sec:results_t1t2}.

\subsubsection{Fixed timing window analysis}\label{sec:analysis_500_sec} 
A 2000~sec window of SK data is selected that occurs within $\pm$1000~sec around the GRB trigger time. 
The inner 1000~sec ($\pm500$sec) is the search window and the outer 1000~sec is used for background estimation. 
This search window was selected by considering the time scale of GRB models and the arrival time delay between gamma-rays and neutrinos caused by neutrino mass. For core-collapse supernovae, the neutrino emission is expected to begin within 1~sec of the explosion, and continue for 
$\sim$10~sec~\cite{supernova1,supernova2}. For neutron star mergers, the neutrino emission is expected to have a duration ranging from a few tens of msec~\cite{gw_nu} to a few sec~\cite{kyutoku} following the merger. In case of the cosmic string model~\cite{lownu2}, the difference between the emission time of gamma-rays and neutrinos is predicted to be less than $\sim$10~sec.

The width of the search window was determined by consideration of several parameters. Relative to photons, neutrinos experience a time-of-flight delay of
\begin{equation}
  \frac{1}{2} \left(\frac{m_\nu}{E_\nu}\right)^2 \times T_\gamma,
\end{equation}
where $m_\nu$ is neutrino mass, $E_\nu$ is neutrino energy and $T_\gamma$ is gamma-ray time of flight. We can estimate $m_\nu$ in this equation from observations of baryon acoustic oscillations, the cosmic microwave background (CMB), and CMB lensing. A combined analysis of these observations can be used to limit the sum of neutrino masses to $\le 0.23$~eV~\cite{Planck}.
By combining these results with those from neutrino oscillation experiments, which measure the difference in the square of the masses, we can determine that the upper limit on the heaviest neutrino is 0.087~eV.
In addition to $m_\nu$, we also need the gamma-ray time-of-flight. To date, the most distant GRB observed is GRB090429B, which has a redshift of 9.4~\cite{GRB090429B} and corresponds to a time-of-flight of $13\times10^9$ years. Using these values together with the lower energy threshold of this search ($E_\nu$ = 8~MeV), we find the maximum neutrino time-of-flight to be delayed by about 24~sec relative to the gamma-ray. This justifies using $\pm$500~sec as a conservative search window.

\begin{figure}[htp]
  \begin{center}
    \includegraphics[width=8cm]{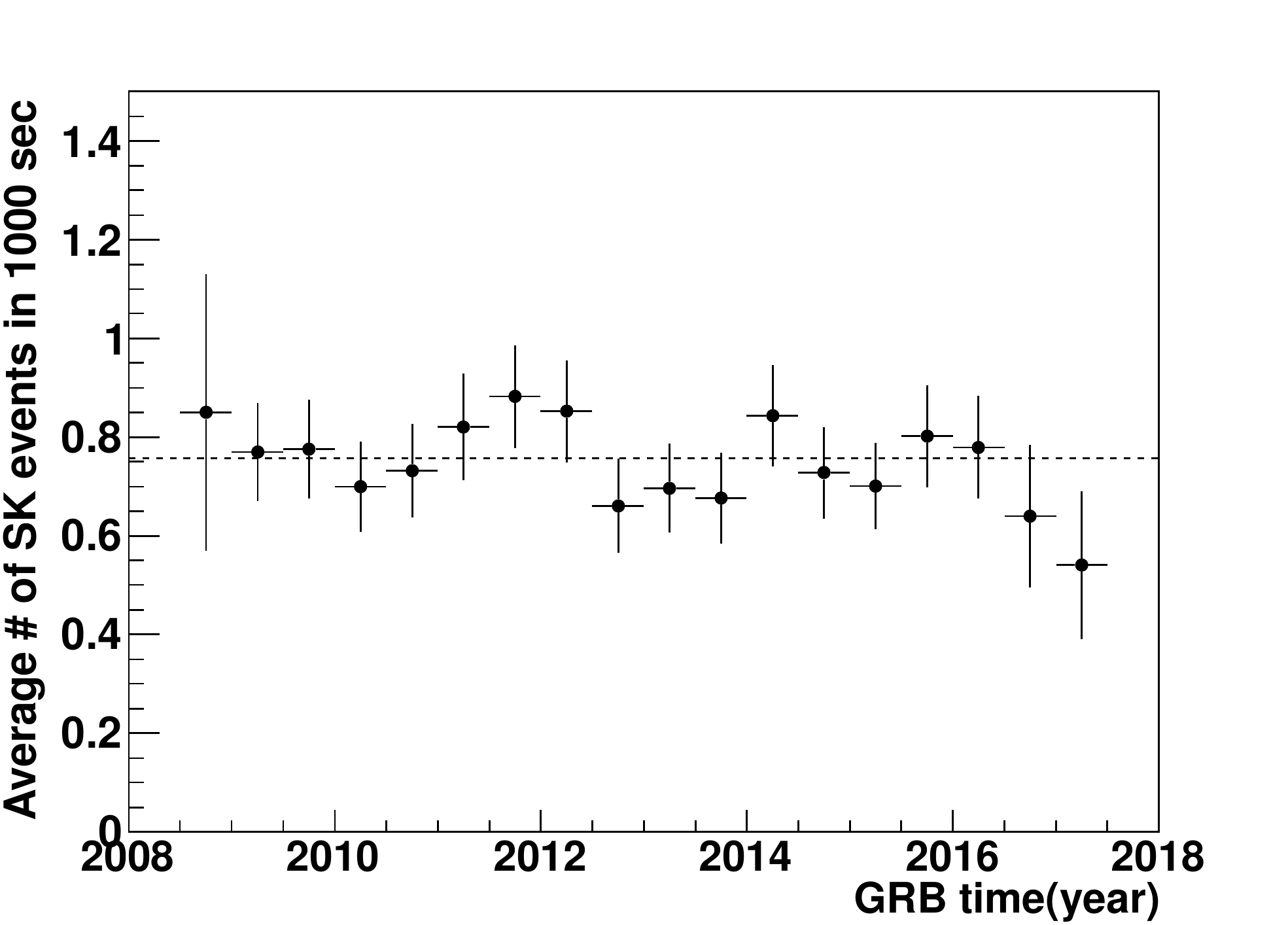}
    \caption{The time variation of the background rate obtained using events in background time window following the first reduction and spallation cut. The horizontal dashed line is the average of whole period.}
    \label{idbg}
  \end{center}
\end{figure}

The outer 1000~sec ($\rm -1000\sim -500~sec$, +500$\sim$+1000~sec) are used for background estimation. The time variation of the background rate is obtained by using events in background time window after the first reduction and spallation cut are applied. These are shown in Fig.~\ref{idbg}. Most of these background events are caused by solar neutrinos, remaining spallation events, and gamma-rays originating from outside of the ID. The background rate is stable in time. Summing over the full sample of 2208 GRBs, we find a total of 251 events remaining in the background window after all reductions. Thus, the average background rate is 0.114~event/1000~sec.

\subsubsection{Variable timing window analysis}\label{sec:analysis_t1t2}
\begin{figure}[htp]

  \begin{center}

    \includegraphics[width=8cm]{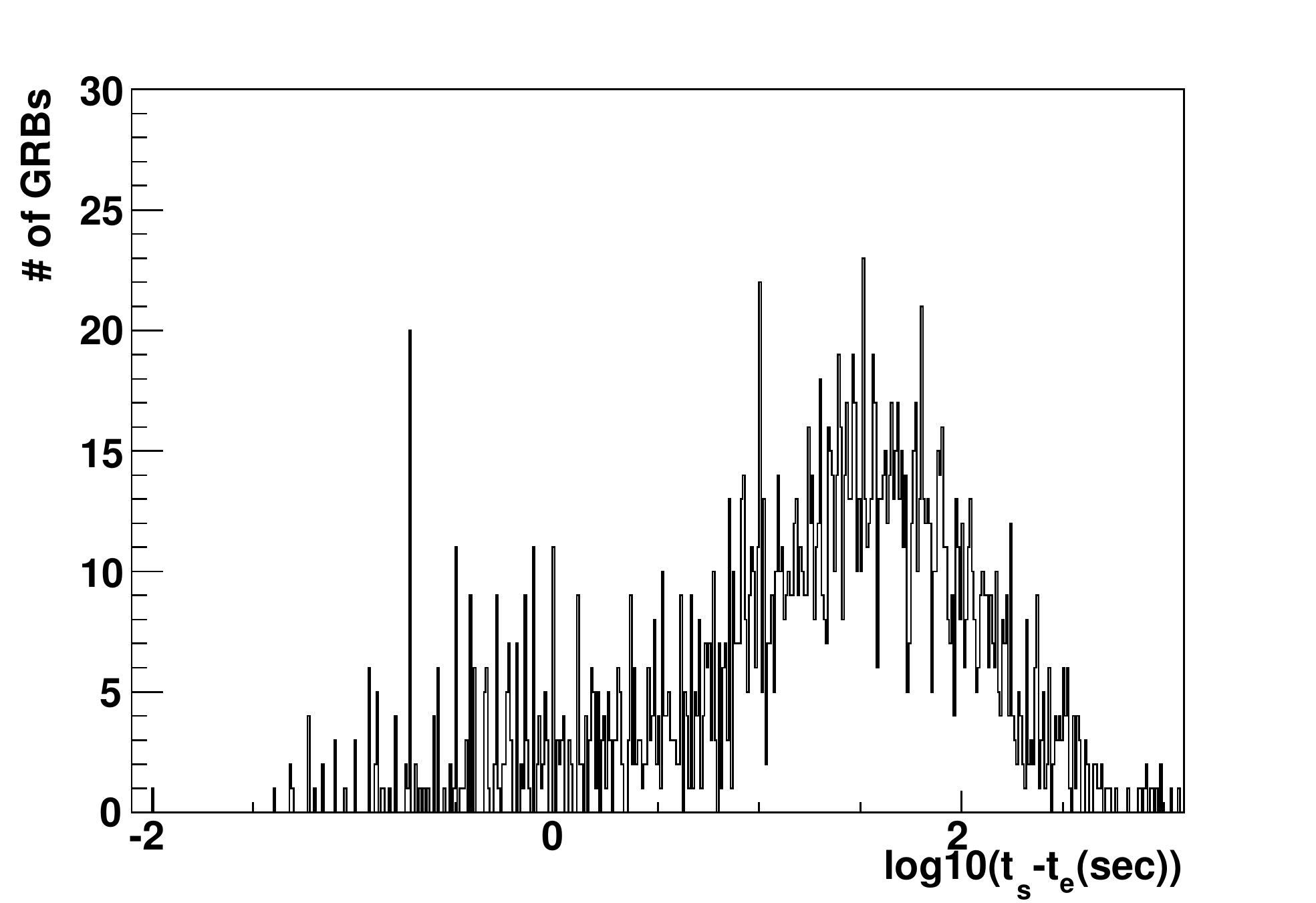}

    \caption{The distribution of GRB duration, $t_e-t_s$, for 2190~GRB events, with timings taken from the GRBweb online catalog~\cite{IClist}. These GRBs are used for the variable time window analysis.}
    \label{t1t2fig}

  \end{center}

\end{figure}

Fig.~\ref{t1t2fig} shows the distribution of GRB duration ($t_e-t_s$) from 0.01~sec to 1000~sec. The variable timing window analysis is designed to reduce the number of background events in short GRBs by reducing the time window of the search. This analysis also increases signal efficiency for very long GRBs, with durations in excess of the fixed window. In this analysis, we count the number of SK events ($N_{ev}$) that are observed between the GRB start time ($t_s$) and end time ($t_e$). The analysis is performed with the 2190~GRBs from the GRBweb online catalog that contain both $t_s$ and $t_e$ information and that have $t_s$ and $t_e$ in $\rm \pm$1000~sec from $t0$. GRB151107A has the maximum $t_e-t_s$, 1202~sec. GRB141102B has the minimum $t_e-t_s$, 0.01~sec.

\subsection{Neutrino search for stacked data}
To enhance the statistical power of this search, a study was performed in which the data from all 2208~GRBs were stacked. In this way, the sensitivity to detect an excess above the background will be improved. 
By accumulating data for many GRBs, only a statistical excess can be observed; even so, this method provides sensitivity for the case where the number of expected events is too small to detect neutrinos in correlation with individual GRBs. This method is called the ``stack analysis". The signal and background windows are defined in the same way as in the fixed timing window analysis (Sec.~\ref{sec:analysis_500_sec}).

\section{Results}
\subsection{Neutrino search for individual GRB}\label{sec:analysis_individual_grb}
\subsubsection{Results of the fixed timing window analysis}\label{sec:results_500_sec}
\begin{figure}[htp]
  \begin{center}
    \includegraphics[width=8cm]{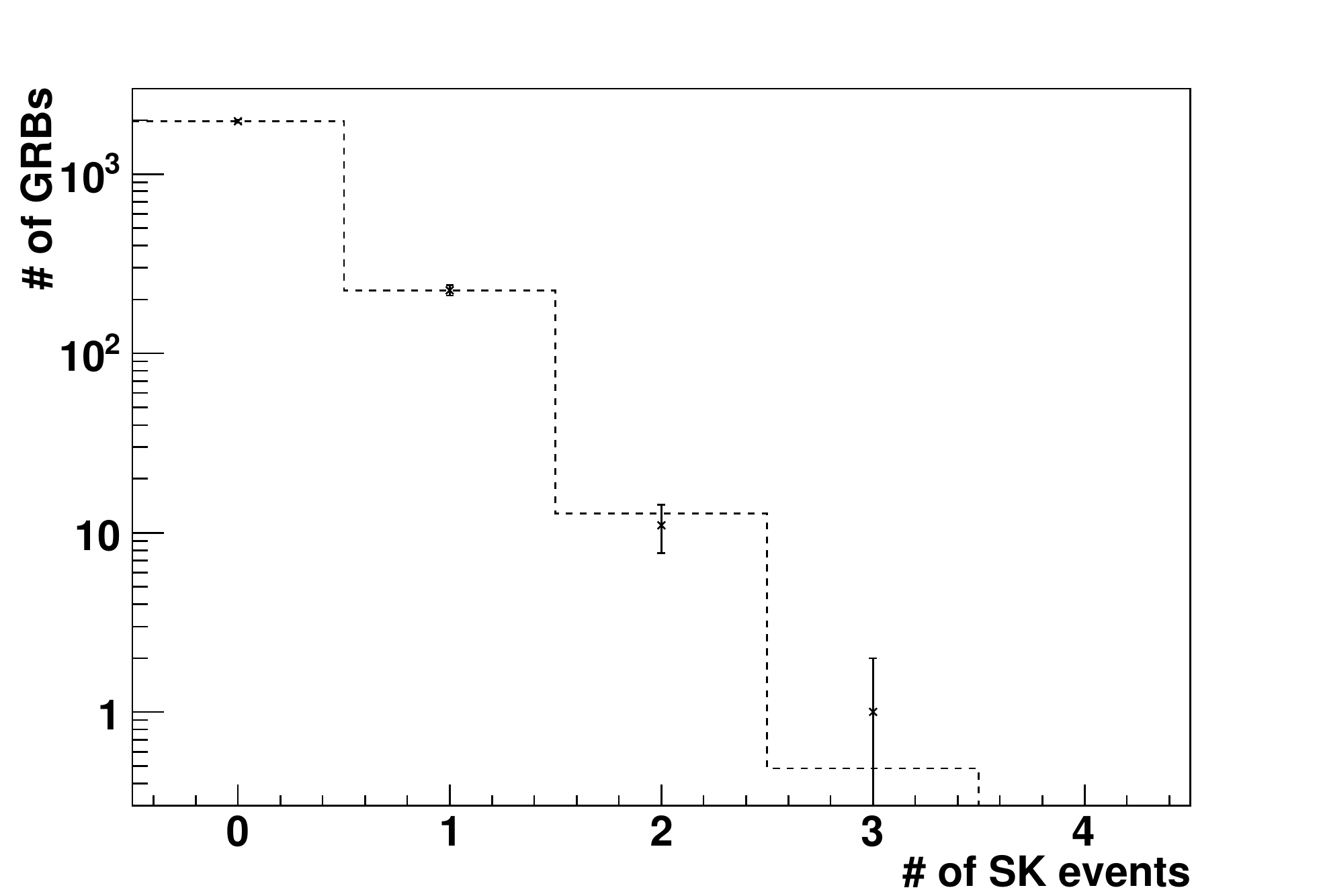}
    \caption{Number of GRBs as a function of SK events detected within the $\pm$500~sec signal window. The dashed line shows the expected number predicted from the estimated background rate. The predicted number of GRBs for 1, 2 and 3 SK events are 224.6, 12.80, 0.486. The points are the observed numbers. The observed number of GRBs for 1, 2 and 3 SK events are 225, 11 and 1.}
    \label{nev}
  \end{center}
\end{figure}

After all reduction cuts are applied, there are 250 events in the signal window and 251 events in the background window. These are total number of events, summed over all 2208 GRBs.
The markers in Fig.~\ref{nev} show the distribution of events in the search window, and the dashed line shows the prediction from a Poisson distribution with an average value of 0.114 and 2208~samples.

\begin{figure}[htp]
  \begin{center}
    \includegraphics[width=8cm]{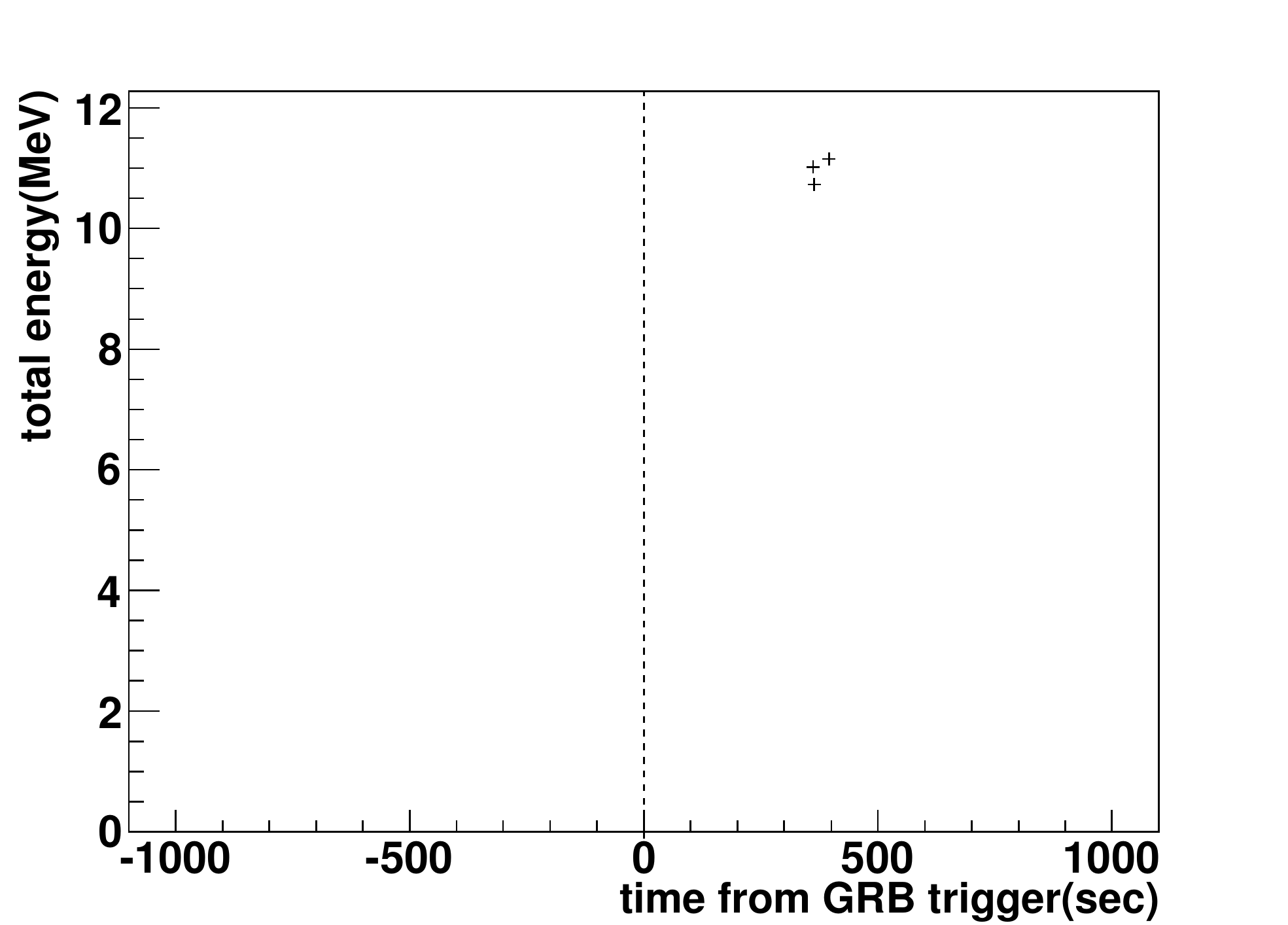}
    \caption{Timing and energy data for SK events within a $\pm$1000~sec window around GRB140616A. The dashed line shows the GRB trigger time ($t0$).}
    \label{fig140616A}
  \end{center}
\end{figure}

In general, the distribution of the observed number of events is consistent with background. The probability of more than 225 GRB with one neutrino event is obtained from the Poisson distribution with an average value of 224.6. The probability is 0.50. Those of more than 11~GRBs with two neutrino events and more than one GRB with three neutrino events are 0.73 and 0.38, respectively. For one GRB (GRB140616A), three neutrino candidate events are found in the fixed search window. Additional checks were performed to determine whether these events constituted a signal. We found that the energy distribution and the transverse distances of these three events are consistent with those of spallation products; thus, we concluded that these events are most likely to be residual spallation events (see Fig.~\ref{fig140616A} and Table~\ref{table140616A}).

\begin{table}[htp]
  \begin{center}
    \begin{tabular}{cccc}
      \hline\\
      Energy (MeV) & Time difference (sec) & Transverse distance (m) & Longitudinal distance (m) \\
      \hline
    11.0 & 0.41 & 2.5 & -6.6 \\
    10.7 & 0.66 & 2.9 & -3.3 \\
     11.2 & 0.44 & 1.9 & -7.5 \\
      \hline
    \end{tabular}
    \caption{Properties of neutrino candidate events remaining in the signal window around GRB140616A after the data reduction is applied. The time difference and spatial distances listed are relative to the most likely parent muon. Transverse distance and longitudinal distance are described in Fig.~\ref{fig:spa_sch}.}
    \label{table140616A}
  \end{center}
\end{table}

\subsubsection{Results of the variable timing window analysis}\label{sec:results_t1t2}
\begin{figure}[htp]
  \begin{center}
    \includegraphics[width=8cm]{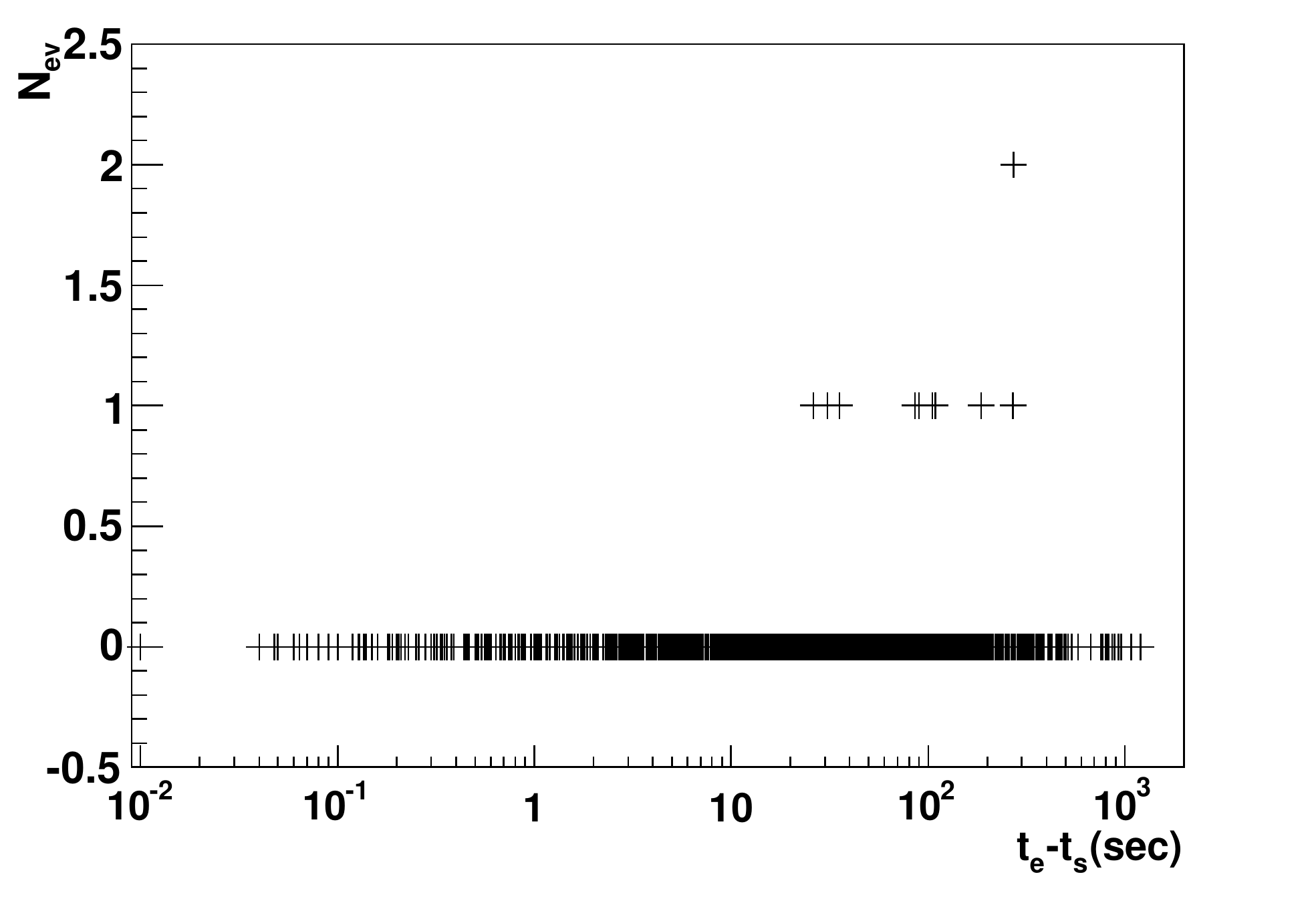}
    \caption{$N_{ev}$ versus $t_e-t_s$ width of data.}
    \label{nevtime}
      \includegraphics[width=8cm]{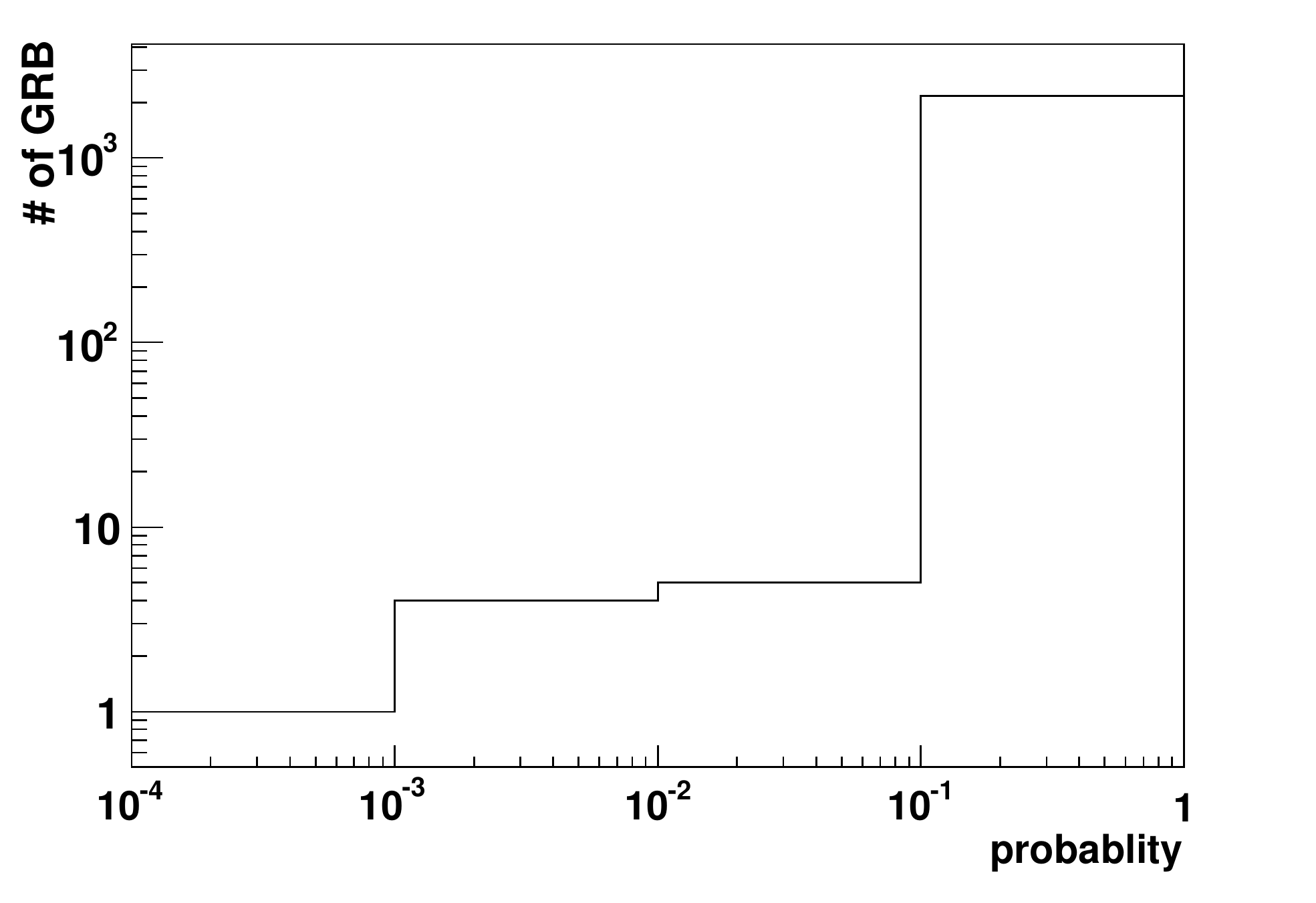}
    \caption{The probability for $N_{ev}$ events to occur randomly in $t_e-t_s$ of data.}
    \label{hprob}
    \end{center}
\end{figure}

The distribution of $N_{ev}$ versus $t_e-t_s$ width is shown in Fig.~\ref{nevtime}. Fig.~\ref{hprob} is a distribution of probability for $N_{ev}$ events would be observed in $t_e-t_s$~sec with a Poisson distribution of the average background rate 0.114~events/1000~sec. The SK data correlated with GRB130315A have the smallest probability of occurring randomly. This GRB is correlated with two SK events that have a 271~sec time separation; the probability for this to occur randomly is 0.00046.

A toy Monte Carlo (MC) simulation was used to check the consistency of the data with the background rate. The estimated number of background events in a particular 
$t_e-t_s$ window is 0.114 events/1000~sec $\times$ ($t_e-t_s$). The probability of detecting $N_{ev}$ events within this window is calculated from a Poisson distribution of this expected background.

To generate each individual MC set, we used the following procedure:

\begin{itemize}
\item For a particular GRB, the number of SK events in a $\pm$1000~sec window, $N_{2000}$, is determined from a Poisson distribution of the background rate, 0.114 events/1000 sec.
\item $N_{2000}$ time differences from the GRB trigger time are randomly allocated in 2000~sec with flat distribution ($t_1, t_2, ..., t_{N_{2000}}$).
\item $t_1, t_2, t_{N_{2000}}$ in the $t_e-t_s$ window is counted. This is the $N_{ev}$ for the simulated GRB.
\item The probability of $N_{ev}$ is calculated as in the data analysis.
\item Repeat the above process for all 2190~GRBs where $t_s$ and $t_e$ data is available.
\item Use the results to make the probability distribution. 
\end{itemize}
Using 10,000~sets of MC generated by this procedure, we were able to evaluate the significance of GRB events with small probability.

\begin{figure}[htp]
  \begin{center}
    \includegraphics[width=10cm]{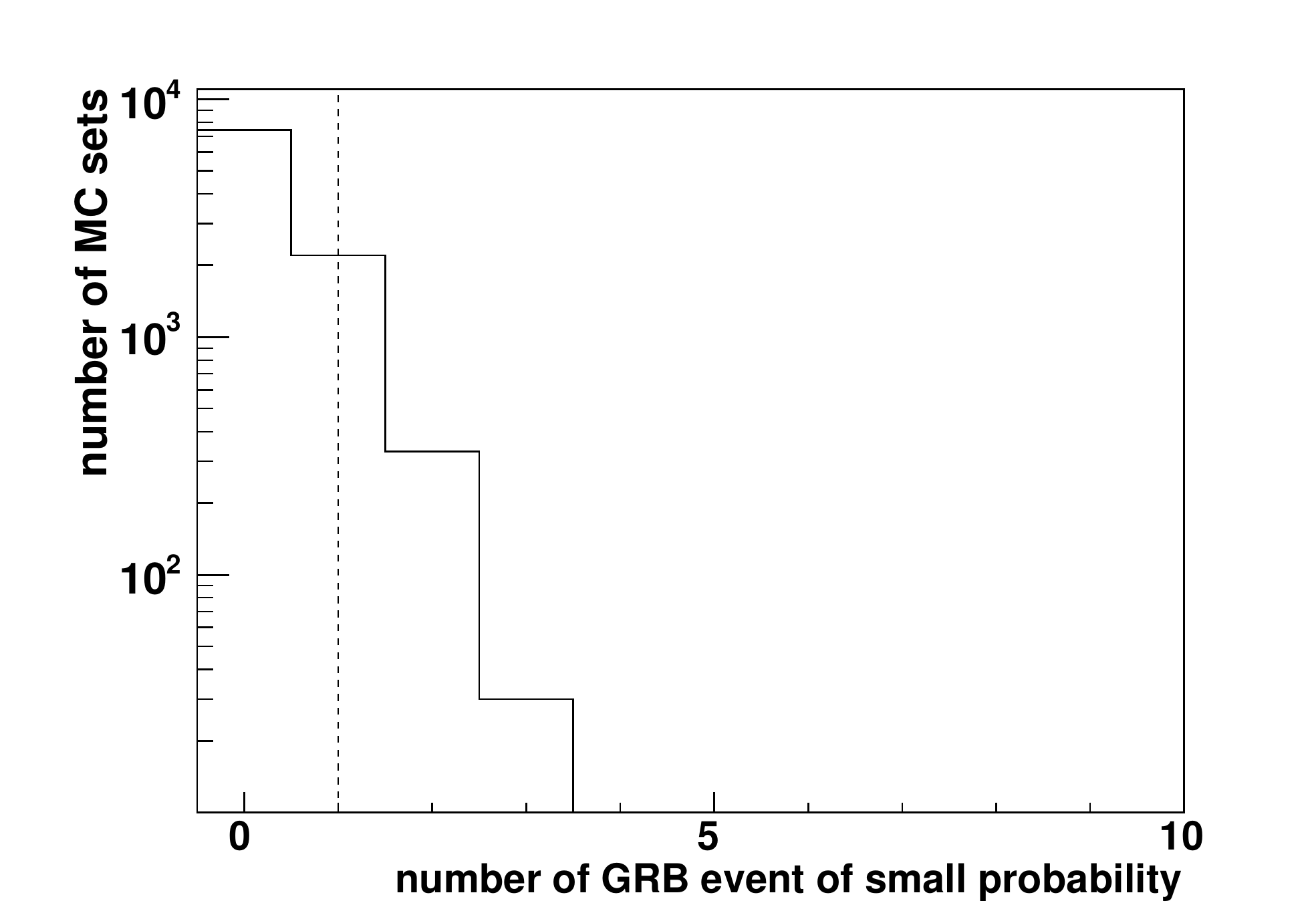}
    \caption{The number of toy MC sets with GRBs that have an $N_{ev}$ corresponding to a probability less than 0.001. The dashed line shows data.}
    \label{toymc1}
  \end{center}
\end{figure}

\begin{figure}[htp]
  \begin{center}
    \includegraphics[width=10cm]{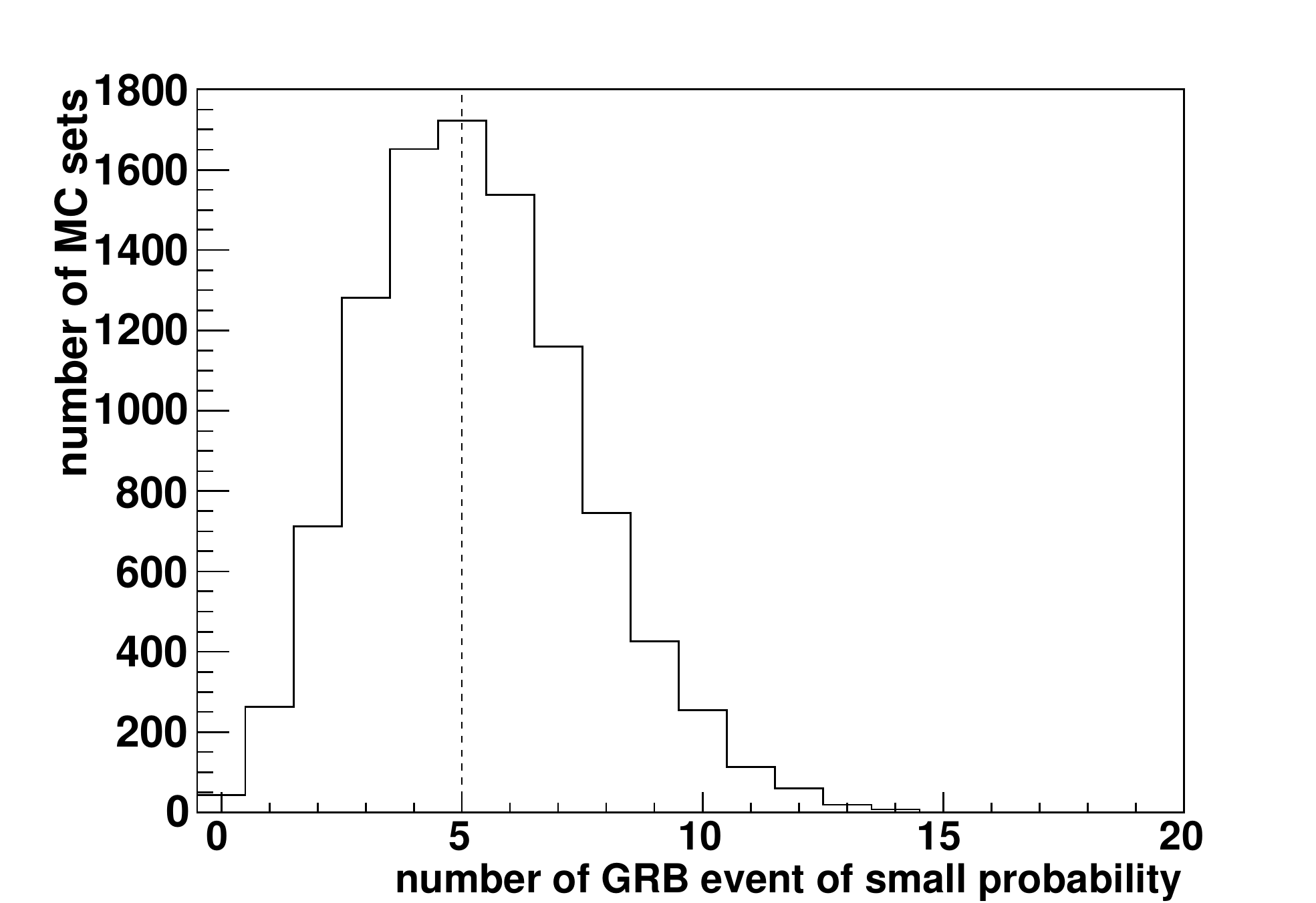}
    \caption{The number of toy MC sets with GRBs that have an $N_{ev}$ corresponding to a probability less than 0.01. The dashed line shows data.}
    \label{toymc2}
  \end{center}
\end{figure}

In the data, 1~GRB has $N_{ev}$ with a corresponding probability of less than 0.001, and 5~GRBs have $N_{ev}$ with a probability less than 0.01. In our simulation, 25.7\% of the toy MC sets have 1 or more GRBs with a probability less than 0.001 (Fig.~\ref{toymc1}), and  60.5\% of these sets have 5 or more GRBs with a probability less than 0.01 (Fig.~\ref{toymc2}). Therefore, we find that the data are statistically consistent with the expected background.


\subsection{Results of the stacked data analysis}\label{sec:analysis_stack}
\begin{figure}[htp]
  \begin{center}
    \includegraphics[width=8cm]{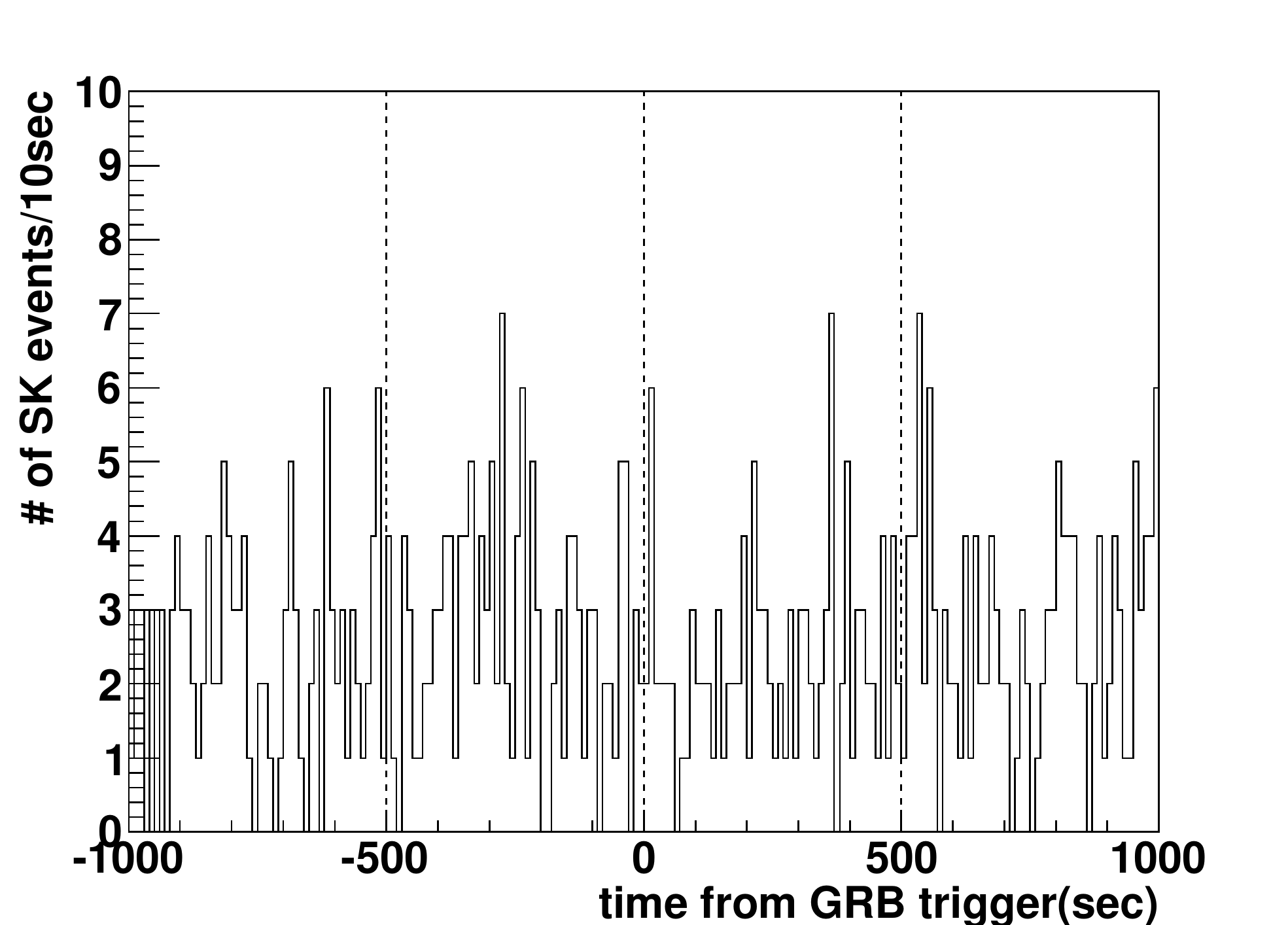}
    \caption{The time distribution of SK events with respect to nearest GRB trigger time, using the full sample of 2208 GRBs. The width of each time bin is 10 seconds. The dashed lines indicate the GRB trigger time and the edges of the signal window.}
    \label{all500.10}
  \end{center}
\end{figure}

\begin{figure}[htp]
  \begin{center}
    \includegraphics[width=8cm]{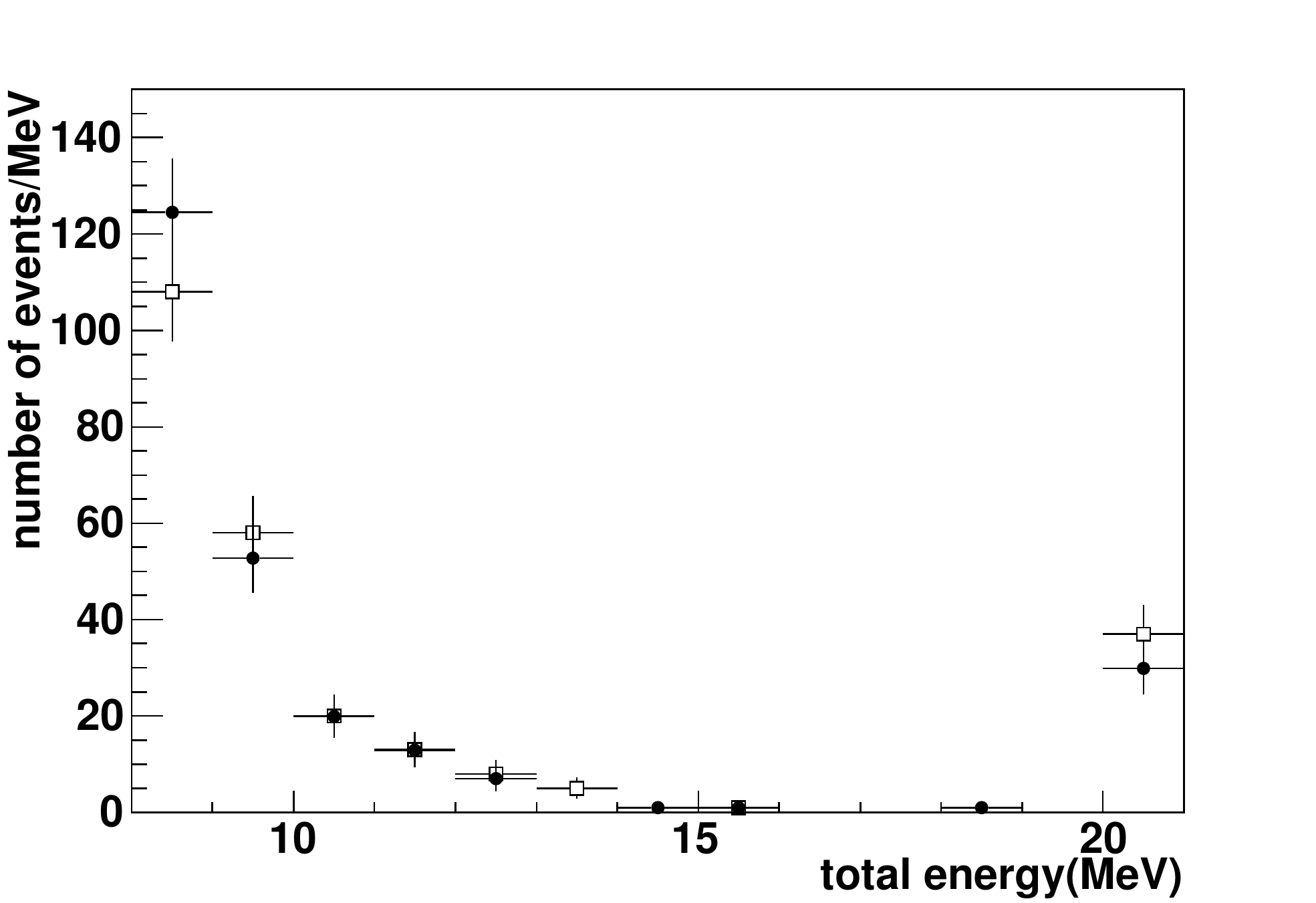}
    \caption{The energy distribution of events in the signal window (open square) and background window (filled circle) from the stacked data analysis. The 21~MeV bin includes all events above 20~MeV.}
    \label{ene}
  \end{center}
\end{figure}

The time distribution of SK events, relative to the GRB trigger time, is shown in Fig.~\ref{all500.10}. This figure contains all SK events that occurred within a $\pm$1000~sec window around any of the 2208 GRB trigger times. No excess of events is seen within the $\pm$500~sec signal window, relative to the background. The energy distribution of events within the signal window was compared to those within the off-time background window. As can be seen in Fig.~\ref{ene}, the two distributions are consistent.


No excess of events was observed, and so we calculated the fluence limit from the stacked data. The stacked data leads to the best expected limit. 
The calculation method is described in~\cite{SK2009}: 
Using a Poisson distribution with the background rate, $N_{90}$, the 90\%~C.L. limit on the number of neutrino events in the signal window can be calculated as
\begin{equation}
\label{eqn:fluence}
\begin{split}
  \int_{N_{bg}}^{N_{bg}+N_{90}} dx \,\rm{Poisson}(\it{N_{obs},x}) &= 0.9 \int_{N_{bg}}^{\infty} dx \,\rm{Poisson}(\it{N_{obs},x}),\\
  \rm{Poisson}(\it{N_{obs},x}) &= \frac{exp(-x) x^{N_{obs}}}{N_{obs}!}
\end{split}
\end{equation}
where $N_{bg}$ is the expected number of background events, $N_{obs}$ is the number of observed events and $\rm{Poisson}(\it{N_{obs},x})$ is the Poisson probability for $N_{obs}$ events with mean of $x$.
When $N_{bg}$ is expected and $N_{obs}$ is observed, the probability that the number of neutrino events is less than $N_{90}$, is 90\%. For the limit calculation, events within an energy window of 8-100~MeV are used. We find $N_{obs}=218$ and $N_{bg}=221$; using these values and equation~\ref{eqn:fluence}, we find the 90\%~C.L. limit $N_{90} = 23.9$.

Once $N_{90}$ has been determined, it can be used to obtain the fluence limit $\Phi$ via this equation:
\begin{equation}
  \Phi = \frac{N_{90}}{N_T\int dE_{\nu} \lambda(E_{\nu}) \sigma(E_{\nu}) \epsilon(E_e)}, \label{limiteq}
\end{equation}
where $N_T$ is the number of target nuclei within the 22.5~kton fiducial volume,
$\lambda$ is the neutrino spectrum normalized to unity,
$\sigma$ is the total neutrino cross section as a function of neutrino energy,
and $\epsilon$ is the detector efficiency as a function of positron energy.
The neutrino spectrum $\lambda$ is assumed to be constant with respect to energy.
The cross-section for inverse beta-decay from Strumia and Vissani~\cite{SVcalc} was used for $\sigma(E_{\nu})$.
The positron total energy is related to the neutrino energy by $E_{\nu} - E_e \simeq 1.293~\rm{MeV}$.
The detector efficiency is calculated as Fig.~\ref{fig:eff} by applying the data reduction process to both the MC sample and the random sample~\cite{Orii_thesis}.
In the energy window between 8~MeV and 100~MeV, the fluence limit per GRB ($\Phi$ / 2208) was found to be $5.07\times10^5$~$\rm{cm}^{-2}$.

\begin{figure}[htp]
  \begin{center}
    \includegraphics[width=8cm]{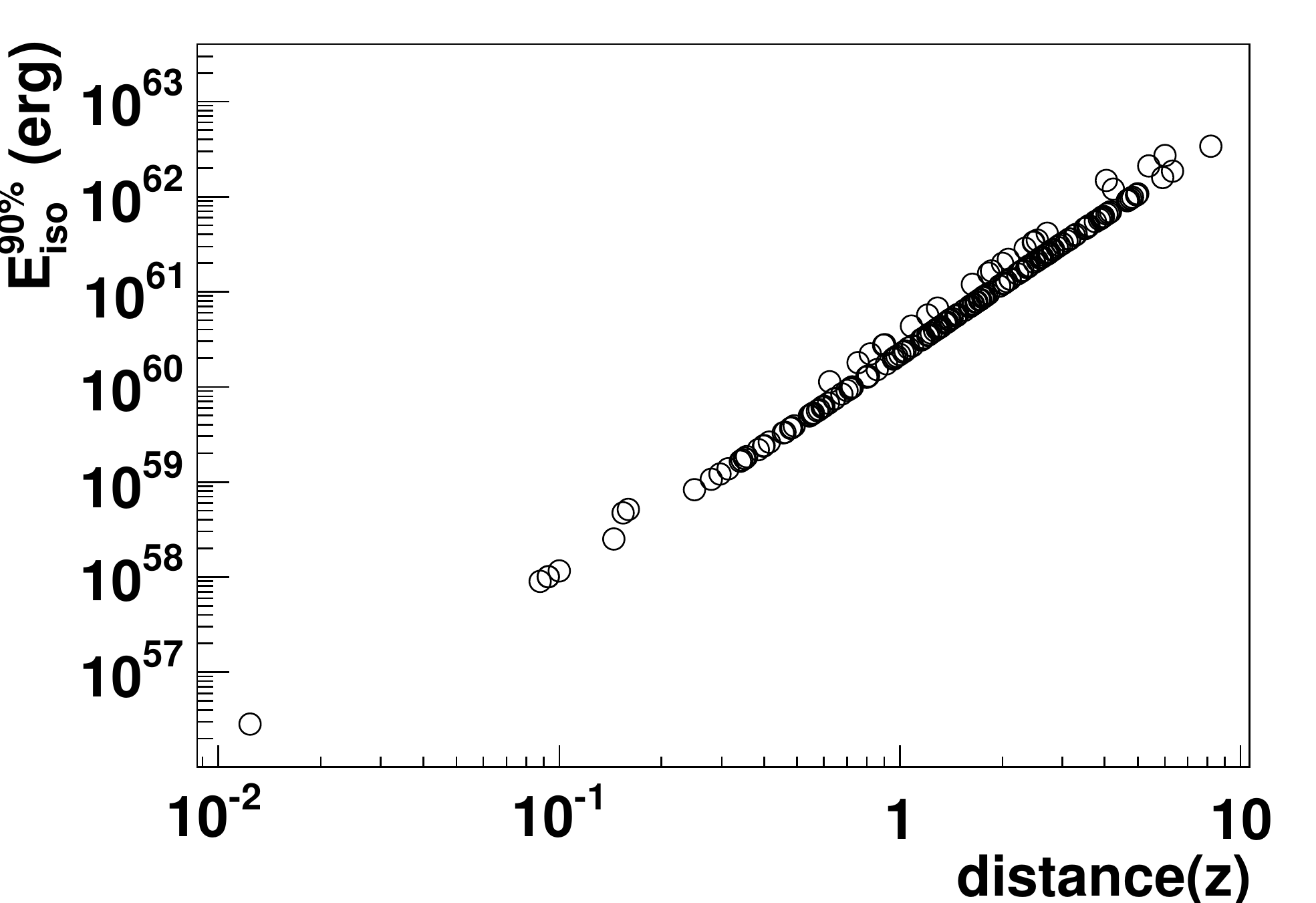}
    \caption{90\% C.L. upper limits on the isotropic energy emitted as neutrinos, plotted as a function of source distance. A flat energy spectrum is assumed in the region from 8~MeV to 100~MeV.}
    \label{lim-dist-flat}
  \end{center}
\end{figure}

\begin{figure}[htp]
  \begin{center}
    \includegraphics[width=8cm]{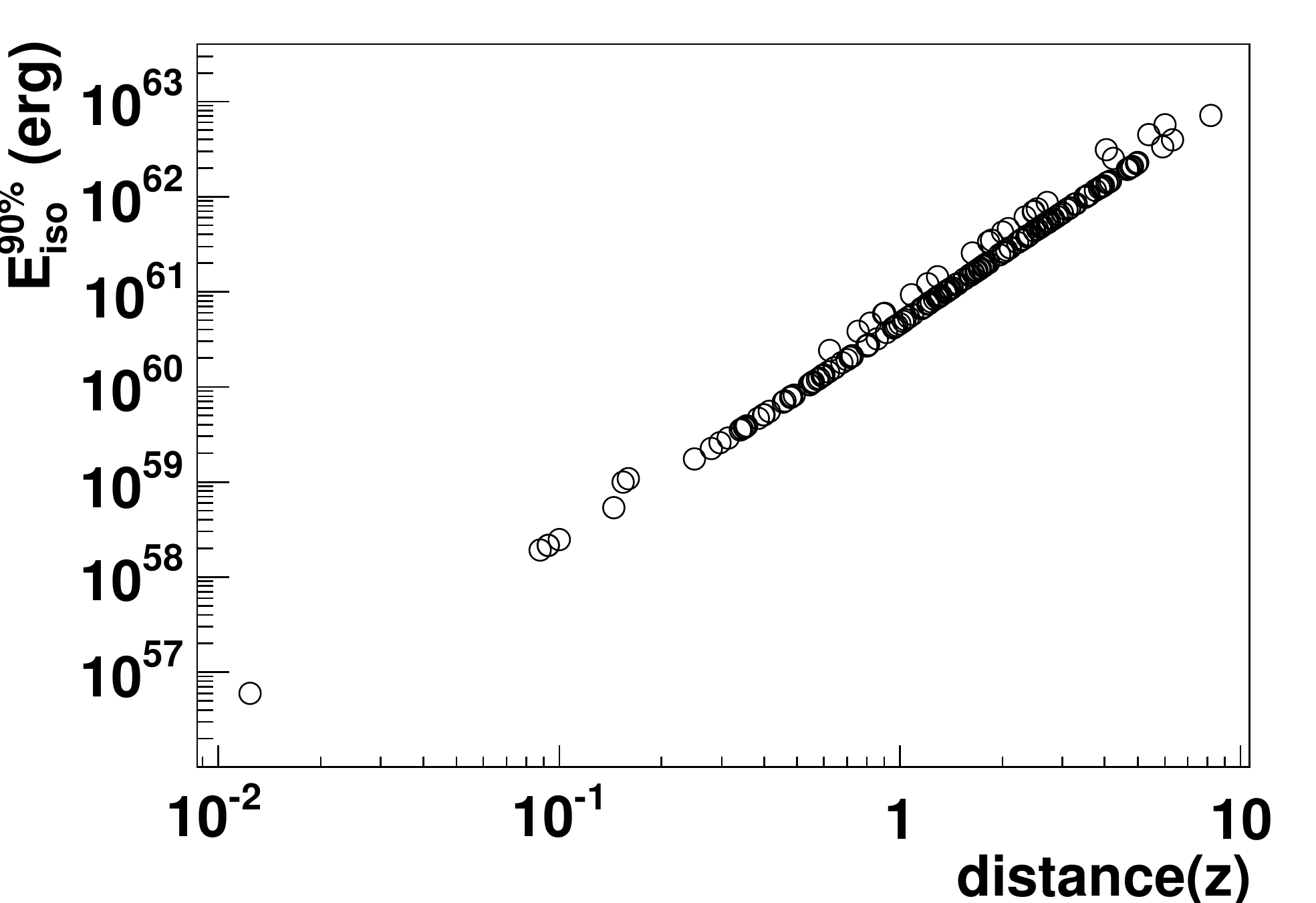}
    \caption{90\% C.L. upper limits on the isotropic energy emitted as neutrinos, plotted as a function of source distance. A Fermi-Dirac spectrum is assumed, with average energy of 20~MeV.}
    \label{lim-dist-fd}
  \end{center}
 \end{figure}

%Limit
For 189~GRBs whose distance from Earth are known, the total energy carried away from the source by neutrinos was calculated. Assuming that the neutrino emission at the source is isotropic, the total energy $E_{iso}$ is
\begin{equation}
  E_{iso} = 4\pi D^2 \int \frac{dn}{dE} \times E dE
\end{equation}
where D is the distance from Earth. Fig.~\ref{lim-dist-flat} shows $E_{iso}$ with the assumption that the energy spectrum is flat from 8~MeV to 100~MeV. Fig.~\ref{lim-dist-fd} shows $E_{iso}$ with the assumption of a Fermi-Dirac spectrum with an average energy of 20~MeV~\cite{lownu4}. The smallest upper limits on $E_{iso}$ are obtained with GRB150906 at z=0.0124~\cite{GRB150906B}; those limits are $2.8\times10^{56}$~erg and $6.0\times10^{56}$~erg for the flat and Fermi-Dirac spectra, respectively.

\begin{figure}[htp]
  \begin{center}
    \includegraphics[width=10cm]{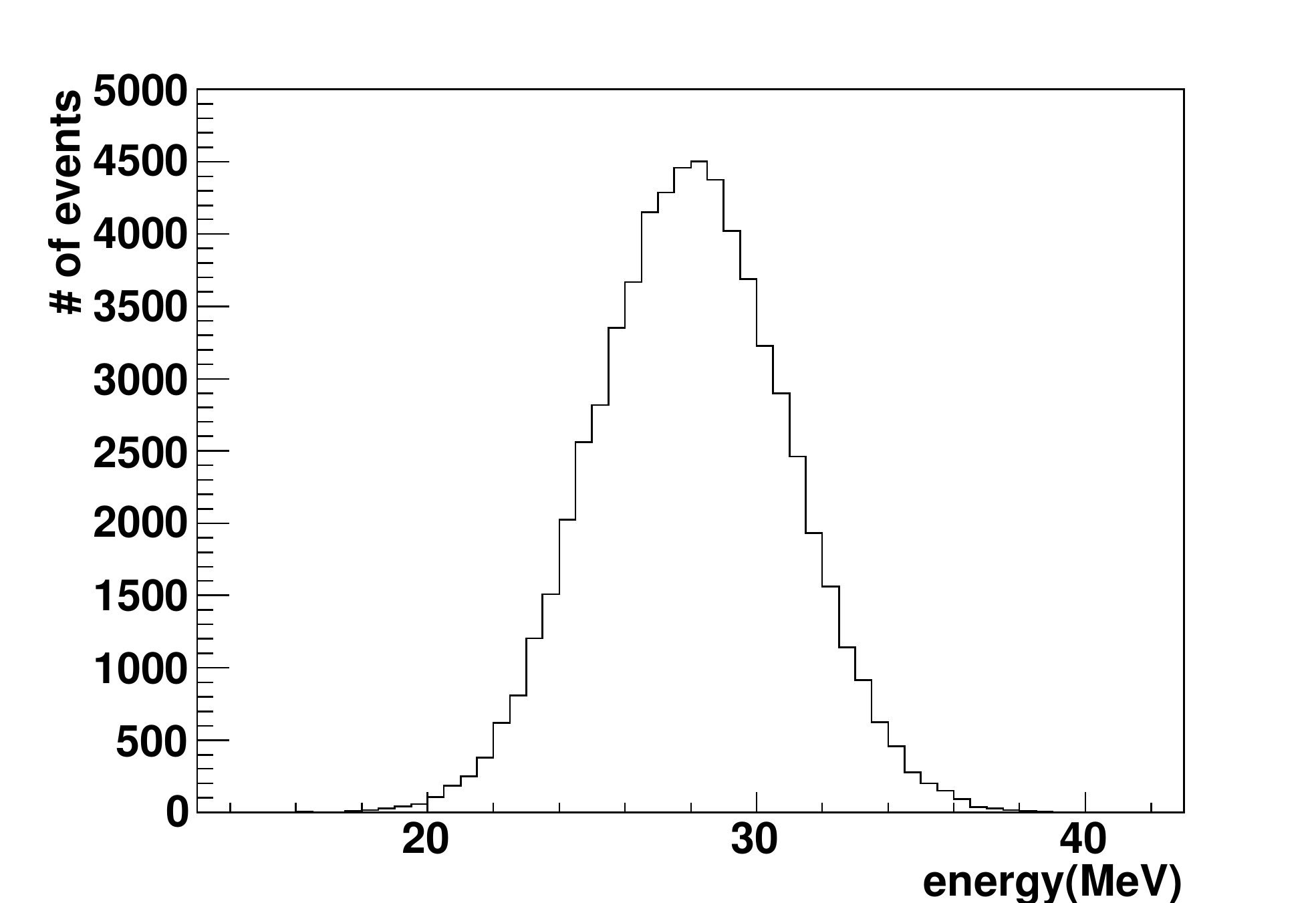}
    \caption{The reconstructed positron energy distribution of the 28~MeV MC sample after reduction.}
    \label{monogaus}
    \end{center}
\end{figure}

We also calculated the fluence limit as a function of energy, using positron energies of 8, 10, 12, 15, 28, 38, 53, 65, and 80~MeV.
Around each target energy, we used events within an energy window to calculate the fluence limit for that energy. The width of that window was determined by using the energy distribution of an MC sample. For each target energy, the reconstructed energy distribution is fitted with a Gaussian distribution to obtain the peak ($E_{peak}$) and the RMS of the energy distribution ($\sigma$). The width of the energy window for that target energy is 3$\sigma$ around $E_{peak}$.
An example can be found in Fig.~\ref{monogaus}, which shows the energy distribution of the 28~MeV MC sample after the full data reduction has been applied. The energy peak 
$E_{peak}$ is at 28.02~MeV, and the deviation is 2.89~MeV; therefore, the energy range for this target energy is 19.35 -- 36.69~MeV.
Table~\ref{enerange} shows the energy peak and the RMS for each MC sample, and Table~\ref{monolim} shows the limits on $N_{90}$ and fluence.

\begin{table}[htp]
  \begin{center}
  \begin{tabular}{cccc}
    \hline
      $E_{gen}$ (MeV) & $E_{peak}$ (MeV) & RMS (MeV) & Total efficiency\\
    \hline
    8 & 7.94 & 1.31 & 0.18\\
    10 & 10.07 & 1.45 & 0.36\\
    12 & 12.10 & 1.60 & 0.42\\
    15 & 15.25 & 1.91 & 0.47\\
    18 & 18.37 & 2.06 & 0.62\\
    21 & 21.23 & 2.23 & 0.76\\
    24 & 24.06 & 2.48 & 0.84\\
    28 & 28.02 & 2.89 & 0.86\\
    38 & 37.97 & 3.68 & 0.85\\
    53 & 52.71 & 4.78 & 0.82\\
    65 & 64.50 & 5.47 & 0.80\\
    80 & 79.38 & 6.07 & 0.77\\
    \hline
  \end{tabular}
  \caption{The relation of MC energy and reconstructed energy. The efficiency is for events in the FV. $E_{gen}$ is generated positron energy, $E_{peak}$ is energy peak after reduction.}
  \label{enerange}
  \end{center}
 \end{table}

Above 38~MeV, it was necessary to enlarge the background window to include the entire data taking period outside of the GRB signal windows ($\pm$1000~sec around $t0$). As shown in Table~\ref{monolim}, the number of expected background events $N_{exp}$ in the usual background window is zero; the enlarged background window is required to obtain an estimate of the background. The live time of the enlarged background window is 2826~days.

The $N_{obs}$ events are not unique to individual energy regions. Due to the finite energy resolution, there is overlap between some regions. For instance, the same four events constitute $N_{obs}$ for the 38~MeV and 53~MeV regions (see Table~\ref{monolim}). Similarly, a single event with energy of 75.4~MeV within the signal window for GRB130509A
comprises the $N_{obs}$ for both the 65~MeV and 80~MeV regions. For all energy regions, $N_{obs}$ is consistent with the expected background.

\begin{figure}[htp]
  \begin{center}
    \includegraphics[width=10cm]{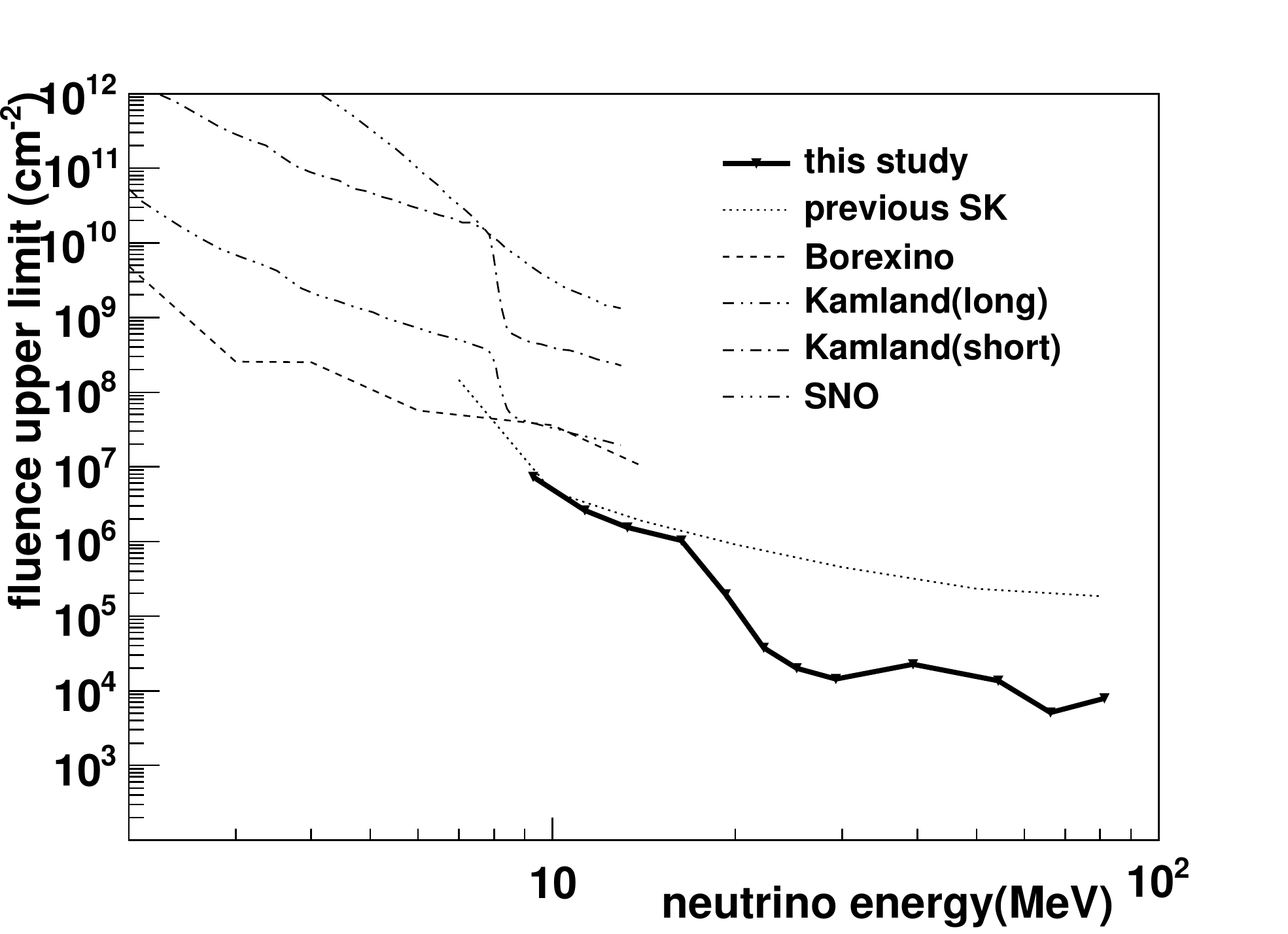}
    \caption{The fluence limit divided by number of GRB. The dotted and solid lines show the results of this work and a previous SK publication~\cite{SK2002}, respectively. The dashed line shows the results from a similar study conducted by Borexino~\cite{Borexino2016}. The dashed double-dotted line and dashed dotted line show the results for long GRB and short GRB by Kamland~\cite{Kamland2015}. The dashed triplicate-dotted line shows the results by SNO~\cite{SNO2014}.}
   \label{compare}
  \end{center}
\end{figure}

The fluence limit using data associated with all GRBs is listed in Table~\ref{monolim}.
Our sample consists of 1813 long GRBs and 323 short GRBs. Fluence limits were calculated for each of these subsamples.
Table~\ref{monolim_long} and Table~\ref{monolim_short} show the number of observed events, the number of expected background events (assuming that the background rate is same for both long and short GRB subsamples), $N_{90}$, and the neutrino fluence limit for long and short GRBs, respectively.
We compare our limits using data associated with all GRBs divided by number of GRB to those obtained from previous studies at SK~\cite{SK2002} and Borexino~\cite{Borexino2016}.
This comparison is shown in Fig.~\ref{compare}. It can be seen that this analysis gives better fluence limits in the energy region of 28-80 MeV. 
In the previous SK study, $N_{90}$ was obtained by counting the number of events in the fixed energy range from 7~MeV to 80~MeV; this was done to get the limit at each energy point. As a result, the $N_{90}$ reported in the previous analysis has a larger value than in the current analysis. This effect is more pronounced in the larger energy region ($>$28~MeV).
The fluence limit at neutrino energy $E'_\nu$ is obtained by replacing $\lambda(E'_\nu)$ in eq.~\ref{limiteq} by a delta function $\delta(E_\nu-E'_\nu)$.

\begin{table}[htp]
\begin{center}
\begin{tabular}{cccccc}
  \hline
  \begin{tabular}{c}
    Positron\\
    total\\
    energy(MeV)
    \end{tabular}&
\begin{tabular}{c}
Reconstructed\\ energy(MeV)
\end{tabular}& $N_{obs}$ & $N_{exp}$ & $N_{90}$ & $\phi$ ($ cm^{-2}$) \\
\hline
8&8.0-11.9&198&210&18.6&7.32$ \times 10^{6}$\\
10&8.0-14.4&212&219&21.5&2.6$ \times 10^{6}$\\
12&8.0-16.9&213&220&21.6&1.54$ \times 10^{6}$\\
15&9.5-21.0&80&67&26&1.03$ \times 10^{6}$\\
18&12.2-24.5&13&10&9.33&1.94$ \times 10^{5}$\\
21&14.6-27.9&1&2&2.99&3.73$ \times 10^{4}$\\
24&16.6-31.5&0&1&2.30&2.00$ \times 10^{4}$\\
28&19.4-36.7&0&0&2.3&1.42$ \times 10^{4}$\\
38&26.9-49.0&4&0(1.78)&6.28&2.26$ \times 10^{4}$\\
53&38.4-67.0&4&0(1.59)&6.44&1.35$ \times 10^{4}$\\
65&48.1-80.9&1&0(1.06)&3.25&5.07$ \times 10^{3}$\\
80&61.2-97.6&1&0(0.9)&3.31&7.84$ \times 10^{3}$\\
\hline \\
\end{tabular}
\caption{The 90~\% C.L. of number of events and the fluence upper limit divided by number of GRB. The number of expected events in the parenthesis is calculated from enlarged background sample (Sec.\ref{sec:analysis_stack}). $N_{obs}$ is number of observed events, $N_{exp}$ is number of expected events, $N_{90}$ is 90~\% C.L. of number of events and $\phi$ is fluence limit.}
\label{monolim}
\end{center}
\end{table}

\begin{table}[htp]
\begin{center}
\begin{tabular}{cccccc}
\hline
\begin{tabular}{c}
Positron\\total \\energy(MeV)
\end{tabular} &
\begin{tabular}{c}
Reconstructed\\ energy(MeV)
\end{tabular}& $N_{obs}$ & $N_{exp}$ & $N_{90}$ & $\phi$ ($ cm^{-2}$) \\
\hline
8&8.0-11.9&152&172&13.2&6.35$ \times 10^{6}$\\
10&8.0-14.4&161&180&14.2&2.09$ \times 10^{6}$\\
12&8.0-16.9&162&181&14.3&1.24$ \times 10^{6}$\\
15&9.5-21.0&59&55&16.6&7.98$ \times 10^{5}$\\
18&12.2-24.5&9&8.2&6.82&1.72$ \times 10^{5}$\\
21&14.6-27.9&1&1.6&3.07&4.67$ \times 10^{4}$\\
24&16.6-31.5&0&0.8&2.30&2.44$ \times 10^{4}$\\
28&19.4-36.7&0&0&2.3&1.73$ \times 10^{4}$\\
38&26.9-49.0&3&0(1.46)&5.32&2.33$ \times 10^{4}$\\
53&38.4-67.0&3&0(1.31)&5.45&1.39$ \times 10^{4}$\\
65&48.1-80.9&1&0(0.87)&3.32&6.32$ \times 10^{3}$\\
80&61.2-97.6&1&0(0.739)&3.38&9.76$ \times 10^{3}$\\
\hline \\
\end{tabular}
\caption{The 90~\% C.L. of number of events and the fluence upper limit for long GRB divided by number of GRB. The number of expected events in the parenthesis is calculated from enlarged background sample (Sec.\ref{sec:analysis_stack}).}
\label{monolim_long}
\end{center}
\end{table}

\begin{table}[htp]
\begin{center}
\begin{tabular}{cccccc}
\hline
\begin{tabular}{c}
Positron\\total \\energy(MeV)
\end{tabular} &
\begin{tabular}{c}
Reconstructed\\energy(MeV)
\end{tabular}& $N_{obs}$ & $N_{exp}$ & $N_{90}$ & $\phi$ ($ cm^{-2}$) \\
\hline
8&8.0-11.9&41&30.7&19.9&5.37$ \times 10^{7}$\\
10&8.0-14.4&45&32&22.9&1.89$ \times 10^{7}$\\
12&8.0-16.9&45&32.2&22.8&1.11$ \times 10^{7}$\\
15&9.5-21.0&18&9.8&15&4.04$ \times 10^{6}$\\
18&12.2-24.5&3&1.46&5.32&7.55$ \times 10^{5}$\\
21&14.6-27.9&0&0.29&2.30&1.96$ \times 10^{5}$\\
24&16.6-31.5&0&0.15&2.30&1.37$ \times 10^{5}$\\
28&19.4-36.7&0&0&2.3&9.73$ \times 10^{4}$\\
38&26.9-49.0&1&0(0.26)&3.67&9.01$ \times 10^{4}$\\
53&38.4-67.0&1&0(0.233)&3.69&5.29$ \times 10^{4}$\\
65&48.1-80.9&0&0(0.155)&2.3&2.46$ \times 10^{4}$\\
80&61.2-97.6&0&0(0.132)&2.3&3.73$ \times 10^{4}$\\
\hline \\
\end{tabular}
\caption{The 90~\% C.L. of number of events and the fluence limit for short GRB divided by number of GRB. The number of expected events in the parenthesis is calculated from enlarged background sample (Sec.\ref{sec:analysis_stack}).}
\label{monolim_short}
\end{center}
\end{table}

\section{Conclusion}
A correlation analysis was used to search for neutrinos associated with GRBs. In this study, SK data from December 2008 to March 2017 was searched for neutrino candidates in coincidence with a set of 2208 GRBs. After applying the event reduction to the SK data sample, 250 candidate events remained in a fixed search window of $\pm$500~sec around each GRB trigger time. This search window corresponds to a total livetime of 25.6 days. The background expected in each 1000~sec search window was estimated to be 0.114~events. No statistically significant excess was observed within the search window, and the energy distribution of the events is consistent with background.  
Three candidate events observed in coincidence with GRB140616A seem to be caused by the spallation of oxygen nuclei by cosmic ray muons.

In addition to the fixed-window analysis, we utilized a variable search window that was determined by the GRB duration time. The number of SK candidate events occurring between the GRB start and end time was used to calculate 
a probability for each GRB. By examining the probability distribution with toy MC, we concluded that the observed number of candidate events in each time window was consistent with expectations from background.

A stacked analysis was performed statistically to search for excess events. This analysis combined the fixed-window data from all 2208~GRBs to create a sample with large statistics. No significant excess was observed in the stacked data, and so we used this sample to calculate the upper limit to the neutrino fluence.
The fluence limit per GRB from 8~MeV to 100~MeV is $5.07\times10^5$~$\rm{cm}^{-2}$.
For 189~GRBs whose distance from Earth are known, the upper limits on the total energy emitted by neutrinos at the source was calculated as a function of the distance to these GRBs.
An energy-dependent fluence limit was also calculated for the full set of GRBs as well as subsets based on GRB duration. The high detection efficiency and large statistics sample used in this study resulted in a world-leading fluence limit between 28~MeV to 80~MeV.

\section*{Acknowledgment}
We gratefully acknowledge the cooperation of the Kamioka Mining and Smelting Company. The Super-Kamiokande experiment has been built and operated from funding by the Japanese Ministry of Education, Culture, Sports, Science and Technology, the U.S. Department of energy, and the U.C. National Science Foundation. Some of us have been supported by funds from the National Research Foundation of Korea NRF-2009-0083526(KNRC) funded by the Ministry of Education (2018R1D1A3B07050696, 2018R1D1A1B07049158), the Japan Society for the Promotion of Science, the National Natural Science Foundation of China under Grants No. 11620101004, the Spanish Ministry of Science, Universities and Innovation (grant PGC2018-099388-B-I00), the National Science and Engineering Research Council (NSERC) of Canada, the Scinet and Westgrid consortia of Compute Canada, the National Science Centre, Poland (2015/18/E/ST2/00758), the Science and Technology Facilities Council (STFC) and GridPPP, UK, the European Union's H2020-MSCA-RISE-2018 JENNIFER2 grant agreement no.822070, and H2020-MSCA-RISE-2019 SK2HK grant agreement no. 872549.
The authors would like to express appreciation to Associate Professor Masaomi Tanaka (Tohoku University) for his kind advice on GRB features.

%Insert the Acknowledgment text here.


\bibliographystyle{ptephy}
\bibliography{bibfile}

\begin{thebibliography}{10}

\bibitem{GRB1973}
R.~W. Klebesadel, I.~B. Strong, and R.~A. Olson, Apj, {\bf 182}, L85 (June
  1973).

\bibitem{Meszaros2006}
P.~Meszaros, Rept. Prog. Phys., {\bf 69}, 2259--2322 (2006),
  {{arXiv:astro-ph/0605208}}.

\bibitem{lownu4}
F.~Halzen and G.~Jaczko, Phys. Rev. D, {\bf 54}, 2779--2783 (Aug 1996).

\bibitem{SK2002}
S.~Fukuda et~al., Apj, {\bf 578}, 317--324 (2002),  {{arXiv:astro-ph/0205304}}.

\bibitem{lownu2}
V.~Berezinsky, B.~Hnatyk, and A.~Vilenkin, Phys. Rev. D, {\bf 64}, 043004 (Jul
  2001).

\bibitem{SK2009}
E.~Thrane $et$ $al.$, Apj, {\bf 697}(1), 730 (2009).

\bibitem{Kamland2015}
K.~Asakura et~al., Apj, {\bf 806}(1), 87 (2015).

\bibitem{Borexino2016}
M.~Agostini $et$ $al.$, Astroparticle Physics, {\bf 86}, 11 -- 17 (2017).

\bibitem{IceCube}
M.~G.~Aartsen $et$ $al.$, Apj, {\bf 843}(2), 112 (jul 2017).

\bibitem{ANITA2011}
A.~G.~Vieregg $et$ $al.$, The Astrophysical Journal, {\bf 736}(1), 50 (2011).

\bibitem{ANTARES2013}
S.~Adrian-Martinez et~al., Astron. Astrophys., {\bf 559}, A9 (2013),
  {{arXiv:1307.0304}}.

\bibitem{SNO2014}
B.~Aharmim $et$ $al.$, Astroparticle Physics, {\bf 55}, 1 -- 7 (2014).

\bibitem{BUST2015}
M.~M. Kochkarov~$et$ $al.$, Physics of Particles and Nuclei, {\bf 46}(2),
  197--200 (Mar 2015).

\bibitem{GCN}
P.~Butterworth S.~D.~Barthelmy, T. L.~Cline, AIP Conf. Proc., {\bf 587}, 213
  (2001).

\bibitem{IClist}
Juan~A. Aguilar (2011),  {{arXiv:1110.5946}}.

\bibitem{SKdetector}
S.~Fukuda $et$ $al.$, Nucl.~Instrum.~Meth.~A, {\bf 501}(2-3), 418--462 (4
  2003).

\bibitem{solar}
K.~Abe,
\newblock Solar neutrino measurements in super-kamiokande-iv (2016),
  {{arXiv:1606.07538}}.

\bibitem{relic2}
K~Bays.~$et$ $al.$, Phys. Rev. D, {\bf 85}, 052007 (Mar 2012).

\bibitem{SVcalc}
Alessandro Strumia and Francesco Vissani, Phys.~Lett.~B, {\bf 564}(1), 42--54
  (2003).

\bibitem{Orii_thesis}
Asato Orii,
\newblock {\em Search for Neutrinos associated with Gamma-ray Bursts in
  Super-Kamiokande},
\newblock PhD thesis, University of Tokyo (3 2019).

\bibitem{supernova1}
K.~Nakazato, K.~Sumiyoshi, H.~Suzuki, T.~Totani, H.~Umeda, and S.~Yamada, Apj
  Supplement Series, {\bf 205}(1), 2 (2013).

\bibitem{supernova2}
T.~Totani, K.~Sato, H.~E. Dalhed, and J.~R. Wilson, Apj, {\bf 496}(1), 216
  (1998).

\bibitem{gw_nu}
K.~Kimuchi, Y.~Sekiguchi, K.~Kyotoku, and M.~Shibata, Class. Quant. Grav., {\bf
  29}, 124003 (2012),  {{arXiv:1206.0509}}.

\bibitem{kyutoku}
Koutarou Kyutoku and Kazumi Kashiyama, Phys. Rev. D, {\bf 97}, 103001 (May
  2018).

\bibitem{Planck}
P.~A.~R. Ade~$et$ $al.$, Astron. Astrophys., {\bf 571}, A16 (2014).

\bibitem{GRB090429B}
A.~Cucchiara $et$ $al.$, Apj, {\bf 736}(1), 7 (Jun 2011).

\bibitem{GRB150906B}
A.~J. {Levan}, N.~R. {Tanvir}, and J.~{Hjorth}, GRB Coordinates Network, {\bf
  18263}, 1 (January 2015).

\end{thebibliography}
%
% once the .bbl file has been generated then place the text in your article.

%\begin{thebibliography}{0}
%
%\end{thebibliography}

%\appendix

%\section{Appendix head}
\end{document}